\documentclass[pra,twocolumn,showpacs,superscriptaddress,floatfix]{revtex4}
\usepackage{epsfig}
\usepackage{amsmath}
\newcommand{\br}{\mathbf{r}}
\newcommand{\bk}{\mathbf{k}}
\newcommand{\brp}{\mathbf{r}_\parallel}
\newcommand{\bkp}{\mathbf{k}_\parallel}
\newcommand{\TE}{{\rm TE}}
\newcommand{\TM}{{\rm TM}}
\newcommand{\kz}{k_{\rm z}}
\newcommand{\kzs}{k_{\rm zl}}
\newcommand{\kzd}{k_{\rm zs}}
\newcommand{\z}{\mathcal{Z}}
\newcommand{\rd}{\hspace{-1mm}{\rm d}}
\begin{document}

\title{Interaction of an atom with layered dielectrics}
\author{Claudia Eberlein}
\author{Robert Zietal}
\affiliation{Department of Physics \& Astronomy,
    University of Sussex,
     Falmer, Brighton BN1 9QH, England}
\date{\today}
\begin{abstract}
We determine the energy-level shift experienced by a neutral atom due
the quantum electromagnetic interaction with a layered dielectric body.
We use the technique of normal-mode expansion to quantize the
electromagnetic field in the presence of a layered, non-dispersive and
non-absorptive dielectric. We explicitly calculate the equal-time
commutation relations between the electric field and vector potential
operators. We show that the commutator can be expressed in terms of a
generalized transverse delta-function and that this is a consequence of
using the generalized Coulomb gauge to quantize the electromagnetic
field. These mathematical tools turn out to be very helpful in the
calculation of the energy-level shift of the atom, which can be in its
ground state or excited. The results for the shift are then analysed
asymptotically in various regions of the system's parameter space -- with a
view to providing quick estimates of the influence of a single
dielectric layer on the Casimir-Polder interaction between an atom and a
dielectric half-space. We also investigate the impact of resonances between
the wavelength of the atomic transition and the thickness of the top
layer.
\end{abstract}

\pacs{31.70.-f, 41.20.Cv, 42.50.Pq}

\maketitle

\section{\label{sec:level1}Introduction}
The question of the interaction between a neutral atom and a macroscopic
dielectric body, once of purely academic interest, has recently been
promoted to a real-life physics problem thanks to the rapid developments
in nanotechnology and experimental techniques. It is no longer the case
that this interaction, the so-called Casimir-Polder interaction, is a
tiny effect that can be ignored in all practical situations. Instead, on
the length-scales that nanotechnology nowadays operates in, dispersion
forces, as they are also called, become significant and may appreciably
influence miniaturized physical systems. Many of the current ambitions
of cold-atom physics towards quantum computation and a variety of
nanotechnological applications involves the trapping and accurate
guiding of single atoms above dielectric substrates, so-called atom
chips. With these the nearby environment of a trapped atom usually
consists of a complicated array of inhomogeneous dielectrics. The
questions then arising are: what are the magnitudes of the
Casimir-Polder forces felt by the atom, and can one possibly engineer
the types and shapes of surrounding materials either to minimize
unwanted dispersion forces or to make them optimally contribute to the
trapping or guiding? In order to investigate such possibilities one
needs to go beyond simple featureless geometries and ground-state atoms
and gain flexibility. The perhaps least sophisticated but still
interesting example to study in this context is to consider a neutral
atom, possibly excited, above a layered dielectric half-space,
cf. Fig. \ref{fig:slab}. If the atom is in its ground state, then the
Casimir-Polder force is always attractive for material surfaces with
refractive indices greater than 1. In such case it is desirable
to derive simple analytical formulae that would allow one to obtain quick
estimates of the magnitudes of the forces involved in terms of the
optical properties of the layer and the substrate \cite{slab}. On the
other hand, if the atom is in its excited state, then, as it is widely
recognised \cite{sipe}, the potential acquires a oscillatory
contribution that can result in a repulsive force. Additionally, the
presence of the layer creates the possibility of a resonance between the
wavelength of the atomic transition and the thickness of the layer,
which could lead to a suppression or enhancement of the interaction.

There exist a variety of theoretical approaches devised to study the
Casimir-Polder interaction (see e.g. \cite{CPReview} for a recent list
of references) but perhaps the most successful ones being the linear
response theory \cite{McLachlan} and phenomenological macroscopic QED
\cite{Buhmann}. By using linear response theory \cite{McLachlan} and
expressing the field susceptibilities in terms of Fresnel reflection
coefficients \cite{sipe,sipe2}, one can express the Casimir-Polder
interaction as an integral along the imaginary frequency axis of the
product of the atomic and field susceptibilities. Thus in practice the
problem is reduced to the calculation of the classical electromagnetic
Green's tensor expressed in terms of Fresnel coefficients. Such
calculations, while straightforward in principle, tend to be quite
tedious and often inevitably lead to the use of numerical
methods. However, there is a benefit to studying problems in quantum
electrodynamics by using physically transparent methods that do not
obscure the basic underlying physics. For the kind of geometry of plane
layered dielectrics considered in this paper, the technique of
electromagnetic field quantization based on a normal-mode expansion
\cite{evan} seems to be best emphasizing the physics of the problem,
namely the fact that the system supports two kinds of modes of the
electromagnetic field \cite{Loudon}: these are travelling modes with a
continuous spectrum and trapped modes with a discrete spectrum,
i.e. occurring at only certain allowed frequencies. The trapped modes
arise because of repeated total internal reflections within the top
layer of higher refractive index than the substrate, and emerge as
evanescent waves outside the wave-guide. This gives rise to an intricate
assortment of evanescent modes outside a layered dielectric where
evanescent waves with continuous spectrum, also arising in a half-space
geometry \cite{evan}, are superposed with discrete evanescent modes that
arise only in the presence of the slab-like waveguide \cite{slab}. In
the framework we apply in this work, in the same spirit as e.g.
\cite{slab,Wu}, the use of standard perturbation theory renders
all calculations explicit and it is possible from the outset to track
down and remove if necessary any ambiguities that tend to remain hidden
in more elaborate theories. For example, linear-response theory results
in an integral over the Fresnel reflection coefficients but gives no
indication of whether the evanescent waves associated with the trapped
modes contribute to the Casimir-Polder interaction or not. The question
is answered at once if the normal-modes approach is used instead, see
\cite{Loudon,Yablonowitch}. Also, interpretations of more complicated
field-theoretical approaches \cite{Bordag} can be put to an explicit
test \cite{slab}.

The purpose of this paper is twofold. Firstly, it aims to support
current experimental efforts by providing a range of analytical formulae
useful for quick estimates of the dispersion forces acting on an atom
placed in the vicinity of the layered dielectric, with particular
emphasis on the corrections caused by the layer as compared to the
standard half-space results reported in \cite{Wu}. It also investigates
the resonant interaction between an excited atom and a layer in the
search for the possible enhancement or suppression of the
Casimir-Polder force. Secondly, it formulates a simple and explicit
theory based on well understood concepts of theoretical physics such as
perturbation theory and electromagnetic field quantization in terms of a
normal-modes expansion. The theoretical aspect, although serving only as
a means to a practical end result, turns out to be interesting in its
own right. The perturbative approach used in this work leads to the
problem of the summation over the modes of the electromagnetic field,
which is non-trivial because of the dual character of the modes of the
electromagnetic field. The task of adding the discrete and continuous
field modes is elegantly accomplished by the use of complex-integration
techniques. This allows us to explicitly show that the canonical
commutation relations between the field operators are satisfied, which
is equivalent to saying that the completeness relation of the
normal-modes holds in the geometry considered. Although this is not a
surprise because the field modes are solutions of a Hermitean operator's
eigenvalue problem, the explicit calculation we carry out provides us
with the mathematics necessary to complete a typical perturbative
calculation in this geometry. It also allows us to cast the end result
in a simple and elegant form that is easy to study analytically in
various asymptotic regimes. The same technique could be applied to any
similar perturbative problem is such a geometry.

This paper is organised as follows. First we quantize the
electromagnetic field in the presence of a layered dielectric, Section
\ref{sec:level2A}. Then, in Section \ref{sec:Comp}, we explicitly prove
the completeness relation for the electromagnetic field modes. Equipped
with the necessary mathematical tools, we proceed to calculate the energy
shift in Section \ref{sec:Shift}, and then study it analytically (Section
\ref{sec:Asym}) and numerically (Section \ref{sec:Num}).

\section{\label{sec:level2A}Field quantization in the presence of a layered boundary}
\begin{figure}[htbp]
\includegraphics[width=8.5 cm, height=6.0 cm]{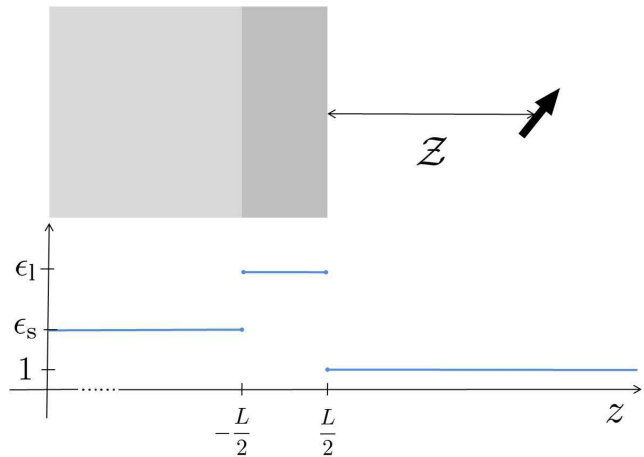}
\caption{\label{fig:slab} Atomic dipole moment in the vicinity of the layered dielectric. The dielectric function is a piecewise constant function of the coordinate $z$.}
\end{figure}

Our ultimate aim is to work out the energy-level shift in an atom caused
by the presence of a layered dielectric. In order to obtain a result
that fully takes into account retardation effects, the quantization of
the electromagnetic field is necessary. To emphasize the physics of the
problem we choose to quantize the electromagnetic field by a normal-mode
expansion as described in \cite{Glauber}.  The dielectric environment we
consider (cf. Fig. \ref{fig:slab}) consists of a substrate, a dielectric
half-space occupying the region of space $z<-L/2$ described by a
dielectric constant $\epsilon_{\rm s}=n_{\rm s}^2$, and on top of that
substrate an additional dielectric layer of thickness $L$, which has a
dielectric constant $\epsilon_{\rm l}=n_{\rm l}^2$. We assume that the
dielectric constant of the layer is higher than that of the substrate
$\epsilon_{\rm l}>\epsilon_{\rm s}$ in order to account for modes that
are trapped inside the layer. Although we work with this assumption, the
final result will turn out to be valid even when the reflectivity of the
substrate exceeds that of the layer, but that is the physically less
interesting case. Throughout this paper we shall assume all dielectric
constants to be frequency independent so that the optical properties of
the system are described solely by a pair of real numbers,
$\epsilon_{\rm l}$ and $\epsilon_{\rm s}$.

To solve Maxwell equations for the electromagnetic field operators in
the Heisenberg's picture we introduce, in the usual manner
\cite{Jackson}, the electromagnetic potentials $\mathbf{A}(\br,t)$ and
$\Phi(\br,t)$ and work in the generalized Coulomb gauge
\begin{equation}
\boldsymbol{\nabla}\cdot[\epsilon(\br)\mathbf{A}(\br)]=0,\label{Coulomb}
\end{equation}
with the dielectric permittivity being a piecewise constant function as
shown in Fig. \ref{fig:slab}. In the absence of free charges one can set
$\Phi(\br,t)=0$ and work only with the vector potential
$\mathbf{A}(\br,t)$ which satisfies the wave equation
\begin{equation}
\nabla^2\mathbf{A}(\br,t)-\epsilon(z)\frac{\partial^2}{\partial t^2}
\mathbf{A}(\br,t)=0,\;\;|z|\neq L/2.\label{Wave}
\end{equation}
Note that right on the interfaces condition (\ref{Coulomb}) is singular
due to discontinuities of the dielectric function and equation
(\ref{Wave}) does not hold at these points. The normal-modes of the field
$\mathbf{f}(\br)e^{i\omega t}$ satisfy the Helmholtz equation
\begin{equation}
\nabla^2\mathbf{f}_{\bk\lambda}(\br)+\epsilon(z)\omega^2
\mathbf{f}_{\bk\lambda}(\br)=0,\;\;|z|\neq L/2,\label{Helm}
\end{equation}
and we have labelled them by their wave-vector $\bk$ and polarization
$\lambda=\{\TE, \TM\}$. This mode decomposition allows one to solve the
field equation (\ref{Wave}) in each distinct region of space separately
and then stitch up the solutions across the interfaces by demanding that
they are consistent with the Maxwell boundary conditions, i.e. that
$\mathbf{E}_\parallel$, $D_\perp$, and $\mathbf{B}$ are all continuous.

The Helmholtz equation (\ref{Helm}) is in fact the eigenvalue problem of
an Hermitean operator\cite{Glauber}
\begin{equation}
\bigg[\frac{1}{\sqrt{\epsilon}}\nabla\times\nabla\times\frac{1}{\sqrt{\epsilon}}\bigg]\sqrt{\epsilon}\mathbf{f}_{\bk\lambda}(\br)=-\omega^2\sqrt{\epsilon}\mathbf{f}_{\bk\lambda}(\br),\label{EigenvalueEq}
\end{equation}
so that we expect the field modes
$\sqrt{\epsilon}\mathbf{f}_{\bk\lambda}(\br)$ to form a complete set of
functions suitable for describing any field configuration. The
completeness relation takes the form
\begin{eqnarray}
\int \rd^2\bkp \sum_{k_z} \hspace{-6mm}\int \;f^i_{\bk\lambda}(\br)\;f^{*j}_{\bk\lambda}(\br')=\delta^\epsilon_{ij}(\br,\br'),\;\;\;z,z'>L/2
\label{comp}
\end{eqnarray}
with $\delta^\epsilon_{ij}(\br,\br')$ being the unit kernel in the
subspace of functions satisfying (\ref{Coulomb}); we shall call this the
generalized transverse delta-function. From quite general considerations
\cite{thesis} we can expect it to be given by
\begin{equation}
\delta^\epsilon_{ij}(\br,\br')=\delta_{ij}\delta^{(3)}(\br-\br')-\nabla_i\nabla'_j\;G(\br,\br')\;\;\label{GenDeltaLayer}
\end{equation}
with the electrostatic Green's function of the Laplace equation given by
\begin{eqnarray}
G(\br,\br')=\frac{1}{4\pi}\frac{1}{|\br-\br'|}-\frac{1}{4\pi}\int_0^\infty\rd k J_0(k\rho)\;e^{-k(z+z')}\nonumber\\
\times\dfrac{\dfrac{n_{\rm l}^2-1}{n_{\rm l}^2+1}-\dfrac{n_{\rm l}^2-n_{\rm s}^2}{n_{\rm s}^2+n_{\rm l}^2}e^{-2kL}}{1-\dfrac{n_{\rm l}^2-1}{n_{\rm l}^2+1}\dfrac{n_{\rm l}^2-n_{\rm s}^2}{n_{\rm s}^2+n_{\rm l}^2}e^{-2kL}}\hspace{.5 cm}\label{GenDeltaLayer1}
\end{eqnarray}
where $\rho=|\brp-\brp'|$ and, for brevity, we have chosen to confine
ourselves to the case $z,z'>L/2$. The function $J_0$ in the above equation is a
Bessel function of the first kind \cite[9.1.1]{AS}. The outline of
the derivation of the Green's function is given in Appendix
\ref{App:Elstatic}.

The sum over all modes in equation (\ref{comp}) is complicated because
the spectrum of the field modes has non-trivial structure. It has been
shown previously \cite{Loudon,Urbach} that the system supports two kinds
of quite distinct types of modes. There are travelling modes going from
left to right or in the opposite direction, and there are guided modes
that are trapped by the dielectric layer, which essentially acts as a
wave-guide. The spectrum of the travelling modes is continuous whereas
the spectrum of the modes trapped in the dielectric layer is discrete
and only some values of the (perpendicular) wave vector are allowed,
namely those satisfying a certain dispersion relation. This dual
character of the spectrum of the field modes is a major obstacle in
working with these modes and calculating, e.g. the energy shift of an
atom nearby, but an elegant solution to this problem has been developed
in \cite{completeness}, whose basic idea we follow here.

We choose the normalization of the mode functions
$\sqrt{\epsilon}\mathbf{f}_{\bk\lambda}(\br)$ according to the convention
\begin{eqnarray}
\int \rd^3\br\;\epsilon(z)\mathbf{f}^*_{\bk\lambda}(\br)\cdot\mathbf{f}_{\bk'\lambda'}(\br)=\hspace{3.5 cm}\nonumber\\
\left\{
\begin{array}{ll}
\delta_{\lambda\lambda'}\delta^{(3)}(\bk-\bk') &\;\; {\rm travelling\;\;modes}\\
\delta_{\lambda\lambda'}\delta^{(2)}(\bkp-\bkp')\delta_{\kz\kz'} &\;\;  {\rm trapped\;\;modes}
\end{array}\right..\;\;\;
\label{normTrapp}
\end{eqnarray}
Then, the electric field $\mathbf{E}(\br)=-\partial_t\mathbf{A}(\br)$
expanded in terms of the normal-modes can be written as
\begin{equation}
\mathbf{E}(\br)=i\sum_\lambda\int \rd^2\bkp \sum_{k_z} \hspace{-6mm}\int\;\;\sqrt{\frac{\omega_\bk}{2\epsilon_0}}a_{\bk\lambda} \mathbf{f}_{\bk\lambda}(\br)e^{-i\omega_\bk t}+{\rm H.C.}
\label{Efield}
\end{equation}
where H.C. stands for Hermitean conjugate. The photon creation and annihilation operators, $a^\dagger_{\bk\lambda}$ and  $a_{\bk\lambda}$, satisfy bosonic commutation relation
\begin{eqnarray}
[a_{\bk\lambda},a_{\bk\lambda'}^\dagger]=\delta_{\lambda\lambda'}
\left\{\begin{array}{l}
\delta^{(3)}(\bk-\bk')\\
\delta^{(2)}(\bkp-\bkp')\delta_{\kz\kz'}
\end{array}\right.\;,\label{Commutators}
\end{eqnarray}
where the top and bottom of the RHS corresponds to the travelling and
trapped photons, respectively. In order to be able to write out the
electromagnetic field operators explicitly one needs to solve the
eigenvalue problem (\ref{Helm}) and determine the spatial dependence of
functions $\mathbf{f}_{\bk\lambda}(\br)$ so we turn our attention to
this now.

\subsection{Travelling modes}\label{sec:trav}
Before we work out the travelling modes, for further convenience, we
introduce Fresnel coefficients for a single interface. For that we
assume that a plane wave is travelling from a medium with refractive
index $n_{\rm b}$ to a medium with the refractive index $n_{\rm a}$, and
that the interface is the $z=0$ plane. Then, the standard Fresnel
reflection and transmission coefficients are given by \cite{Jackson}
\begin{eqnarray}
r_{\TE}^{\rm{ba}}\;&=\;&\frac{k_{\rm zb}-k_{\rm za}}{k_{\rm zb}+k_{\rm za}}\;,\;\;\;\;\;\;\;\;\;\;\;\;\;t^{\rm ba}_{\TE}=\frac{2k_{\rm zb}}{k_{\rm zb}+k_{\rm za}}\;, \label{Fresnel}\\ r_{\TM}^{\rm{ba}}\;&=\;&\frac{k_{\rm zb}/n_{\rm b}^2-k_{\rm za}/n_{\rm a}^2}{k_{\rm zb}/n_{\rm b}^2+k_{\rm za}/n_{\rm a}^2}\;,\;t^{\rm ba}_{\TM}=\frac{2k_{\rm zb}/n_{\rm a}n_{\rm b}}{k_{\rm zb}/n_{\rm b}^2+k_{\rm za}/n_{\rm a}^2}\;,\nonumber
\end{eqnarray}
where $k_{\rm zi}$ are the components of the wave vectors perpendicular to the interface in the medium $\rm{i}=\{\rm{a},\rm{b}\}$. 

The geometry of the problem (cf. Fig. \ref{fig:slab}) naturally divides the space into three distinct regions. Consequently there are three wave vectors to be distinguished. The wave vector in vacuum ($z>L/2$)
\begin{equation}
\bk^\pm=( k_{\rm x},k_{\rm y},\pm k_{\rm z} )=(\bkp,\pm k_{\rm z} ),
\end{equation}
the wave vector in the dielectric layer ($|z|<L/2$)
\begin{eqnarray}
\bk_{\rm l}^\pm =(k_{\rm x}, k_{\rm y},\pm k_{\rm zl})=(\bkp,\pm k_{\rm zl}),
\end{eqnarray}
and the wavevector in the substrate ($z<-L/2$)
\begin{eqnarray}
\bk_{\rm s}^\pm =(k_{\rm x}, k_{\rm y},\pm k_{\rm zs})=(\bkp,\pm k_{\rm zs}).
\end{eqnarray}
The components of the wave vector that are parallel to the surface are the same for all three regions of space. This follows directly from the requirement that the boundary conditions  must be satisfied at all points of a given surface i.e. the spatial phase factors $e^{i\bk_i\cdot\br}$ must be equal at $z=\pm L/2$ for all $\brp$. The different signs of the $z$-components of the wave vectors correspond to the waves propagating in different directions. However, the direction of the propagation of a particular mode needs to be consistent in all three layers so we require that on the real axis
\begin{equation} 
\rm{sign}(k_{\rm z})=\rm{sign}(k_{\rm zl})=\rm{sign}(k_{\rm zs}).
\end{equation}
Since the frequency $\omega$ of a single mode is fixed, the $z$-components of the wave vectors in the dielectric are related to the vacuum wave vector $k_{\rm z}$ by
\begin{eqnarray}
k_{\rm zl}\;&=\;&\sqrt{(n_{\rm l}^2-1)\bkp^2+n_{\rm l}^2k_{\rm z}^2}\;,\label{kzsAskz}\\
k_{\rm zs}\;&=\;&\sqrt{(n_{\rm s}^2-1)\bkp^2+n_{\rm s}^2k_{\rm z}^2}\;.\label{kzdAskz}
\end{eqnarray}
The mode functions $\mathbf{f}_{\bk\lambda}(\br)$ are transverse everywhere except right on the interfaces $z=\pm L/2$, cf. (\ref{Coulomb}). To ensure this transversality, it is convenient to introduce orthonormal polarisation vectors 
\begin{equation}
\mathbf{f}_{\bk\lambda}(\br)=\hat{\mathbf{e}}_\lambda(\bk)f_{\bk\lambda}(\br)
\end{equation}
defined as
\begin{eqnarray}
\hat{\mathbf{e}}_{\TE}(\mathbf{\boldsymbol{\nabla}})\;&=\;&(-\Delta_\parallel)^{-1/2}\big(-i\nabla_{\rm y},i\nabla_{\rm x},0\big),\nonumber\\
\hat{\mathbf{e}}_{\TM}(\mathbf{\boldsymbol{\nabla}})\;&=\;&(\Delta_\parallel\Delta)^{-1/2}\big(-\nabla_{\rm x}\nabla_{\rm z},-\nabla_{\rm y}\nabla_{\rm z},\Delta_\parallel\big),\;\;\;\label{TM}
\end{eqnarray}
with $\Delta$ being the Laplace operator expressed in Cartesian
coordinates and it is understood that the above operators act on the
factors of the type $e^{i\bk^\pm_{\rm i}\br}$,
i.e. ${\hat{\mathbf{e}}_{\lambda}(\mathbf{\bk}_{\rm
i}^\pm)\equiv\hat{\mathbf{e}}_{\lambda}(\mathbf{\nabla})e^{i\bk^\pm_{\rm
i}\br}}$. Polarization vectors defined in such a way are normalized to
unity provided all three components of the wave vector are real. However,
they are not of unit length in the case of evanescent waves which have
wave vectors with pure imaginary components. The spatial dependence of
the mode functions is worked out requiring that each mode consists of
the incoming, reflected and transmitted parts that are joined together
by standard boundary conditions across the interfaces, i.e. that
$\mathbf{E}_\parallel$, $D_\perp$ and $\mathbf{B}$ are continuous. From
this it is straightforward to derive that the travelling modes of the
system incident from the left, normalized according to
(\ref{normTrapp}), are given by
\begin{eqnarray}
\mathbf{f}^L_{\bk\lambda}(\br)=\frac{\hat{\mathbf{e}}_{\lambda}(\mathbf{\nabla})}{(2\pi)^{\frac{3}{2}}n_{\rm s}}
\left\{
\begin{array}{lr}
e^{i\bk_{\rm s}^+\cdot\br} +R^L_\lambda e^{i\bk_{\rm s}^-\cdot\br} & z < -L/2\\
I^L_\lambda e^{i\bk_{\rm l}^+\cdot\br} +J^L_\lambda e^{i\bk_{\rm l}^-\cdot\br} & |z|<L/2\\
T^L_\lambda e^{i\bk^+\cdot\br} & z > L/2
\end{array}\right. \;,\nonumber\\\label{LeftIncident}
\end{eqnarray}
whereas the right-incident modes are given by
\begin{eqnarray}
\mathbf{f}^R_{\bk\lambda}(\br)=\frac{\hat{\mathbf{e}}_{\lambda}(\mathbf{\nabla})}{(2\pi)^{\frac{3}{2}}}
\left\{
\begin{array}{lr}
T^R_\lambda e^{i\bk_{\rm s}^-\cdot\br}  & z < -L/2\\
I^R_\lambda e^{i\bk_{\rm l}^-\cdot\br} +J^R_\lambda e^{i\bk_{\rm l}^+\cdot\br} & |z| <  L/2\\
e^{i\bk^-\cdot\br} +R^R_\lambda e^{i\bk^+\cdot\br} & z >  L/2
\end{array}\right.\; . \nonumber\\\label{RightIncident}
\end{eqnarray}
For the sake of clarity the complete list of reflection and transmission
coefficients is given in Appendix \ref{App:Fresnels}. Here we only write
down the ones most relevant for the calculation at hand:
\begin{eqnarray}
R_\lambda^R\;&=\;&\frac{r_\lambda^{\rm vl}+r_\lambda^{\rm ls}e^{2ik_{\rm zl}L}}{1+r_\lambda^{\rm vl}r_\lambda^{\rm ls}e^{2ik_{\rm zl}L}}e^{-ik_{\rm z}L},\label{R}\\
T_\lambda^L\;&=\;&\frac{t_\lambda^{\rm sl}t_\lambda^{\rm lv}e^{(2ik_{\rm zl}-ik_{\rm zs}-ik_{\rm z})L/2}}{1+r_\lambda^{\rm sl}r_\lambda^{\rm lv}e^{2ik_{\rm zl}L}}.
\end{eqnarray}

\subsection{Trapped modes}\label{sec:trap}
Trapped modes arise from repeated total internal reflections within the layer of higher refractive index $n_{\rm l}$. This happens when the angle of incidence of the incoming wave is sufficiently high and exceeds the critical angle. This critical angle is different for the two opposite waveguide interfaces. First consider the layer-vacuum interface. From equation (\ref{kzsAskz}) we can obtain the reciprocal relation expressing the $k_{\rm z}$ in terms of the $k_{\rm zl}$
\begin{equation}
k_{\rm z}=\frac{1}{n_{\rm l}}\sqrt{k_{\rm zl}^2-(n_{\rm l}^2-1)\bkp^2}\; .
\end{equation}
Thus, whenever $k^2_{\rm zl}<(n_{\rm l}^2-1)\bkp^2$ then $k_{\rm z}$ becomes pure imaginary
\begin{equation}
k_{\rm z}=+\frac{i}{n_{\rm l}}\sqrt{(n_{\rm l}^2-1)\bkp^2-k_{\rm zl}^2}\; ,
\label{eqn:ImaginaryKz}
\end{equation}
and we have a mode that exhibits evanescent behaviour on the vacuum
side. The sign of the square root is chosen such that these modes decay
exponentially when one goes away from the layer in the positive
$z$-direction. This also ensures that there truly is total internal
reflection, i.e. that $|r_\lambda^{\rm vl}|^2=1$.

However, since on the other side of the waveguide we have a substrate
rather than vacuum, not all of the modes that get totally internally
reflected at the vacuum-layer interface necessarily get trapped. From the relation 
\begin{equation}
k_{\rm zs}=\frac{n_{\rm s}}{n_{\rm l}}\sqrt{k^2_{\rm zl}-\bkp^2\bigg(\frac{n^2_{\rm l}}{n^2_{\rm s}}-1\bigg)}
\end{equation}
we obtain the condition of total internal reflection for the substrate-layer interface to be $k^2_{\rm zl}\leq(n^2_{\rm l}/n^2_{\rm s}-1)\bkp^2$. Therefore, modes satisfying the condition 
\begin{equation}
(n^2_{\rm l}/n^2_{\rm s}-1)\bkp^2\leq k_{\rm zl}^2\leq (n_{\rm l}^2-1)\bkp^2
\label{evanContCond}
\end{equation}
are not trapped but appear in vacuum as a continuous spectrum of
evanescent waves that are accounted for among the left-incident
travelling modes. (They are analogous to the evanescent modes that occur
at a single-interface half-space, for which the normal-mode quantization
was first presented in \cite{evan}.) On the other hand, trapped modes occur if
\begin{equation}
0\leq k_{\rm zl}^2\leq(n^2_{\rm l}/n^2_{\rm s}-1)\bkp^2.
\label{trappedCond}
\end{equation}
The procedure for obtaining the trapped modes is largely equivalent to
that of the travelling modes. They can be written in the form
\begin{eqnarray}
\mathbf{f}^T_{\bk\lambda}(\br)=N_\lambda \hat{\mathbf{e}}_{\lambda}(\mathbf{\nabla})
\left\{
\begin{array}{lr}
T^{\rm ls}_\lambda e^{i\bk_{\rm s}^-\cdot\br}  & z<-L/2\\
V_\lambda e^{i\bk_{\rm l}^-\cdot\br} + e^{i\bk_{\rm l}^+\cdot\br} & |z|< L/2\\
T^{\rm lv}_\lambda e^{i\bk^+\cdot\br} & z > L/2
\end{array}\right. .\nonumber\\
\end{eqnarray}
The boundary conditions are imposed on both interfaces. From the
boundary at $z=-L/2$ we get
\begin{eqnarray}
T^{\rm ls}_\lambda\;&=\;&(t_\lambda^{\rm ls}/r_\lambda^{\rm ls})e^{-i(k_{\rm zl}+\kzd)L/2}\;,\nonumber\\
V_\lambda\;&=\;& (1/r_\lambda^{\rm ls})e^{-ik_{\rm zl}L}\label{V1}\;,
\end{eqnarray}
whereas from the $z=L/2$ boundary 
\begin{eqnarray}
T^{\rm lv}_\lambda\;&=\;& t_\lambda^{\rm lv}e^{-i(k_{\rm zl}-\kz)L/2}\;,\nonumber\\
V_\lambda\;&=\;& r_\lambda^{\rm lv}e^{ik_{\rm zl}L}.\label{V2}
\end{eqnarray}
Since both equations, (\ref{V1}) and (\ref{V2}), need to be
simultaneously satisfied we obtain a dispersion relation for these modes,
\begin{equation}
1+r_\lambda^{\rm vl}r_\lambda^{\rm ls}e^{2ik_{\rm zl}L}=0\; , 
\label{disp}
\end{equation}
which determines the allowed values of $\kzs$ within the layer. Since we
will be dealing with an atom on the vacuum side it will be necessary to
express the dispersion relation in terms of $k_z$ rather than
$\kzs$. It is straightforward to show that the allowed values of the $z$-component of the evanescent waves' wave vector appearing on the vacuum side are given by numbers $q^n_{\lambda}$:
\begin{eqnarray}
q^n_{\TE} \;&=\;& \left\{ \kz : \kz+i\kzs(\kz) \tan [\phi_{\TE}(\kz)]=0 \right\}, \nonumber\\
q^n_{\TM} \;&=\;& \left\{ \kz : \kz+i\kzs(\kz)/n_{\rm l}^2 \tan [\phi_{\TM}(\kz)]=0 \right\},\nonumber\\
\label{disp2}
\end{eqnarray}
with
\begin{eqnarray}
\phi_{\TE}(\kz)\;&=\;& \arg\left[(\kzs+\kzd)e^{-i\kzs L}\right],\nonumber\\
\phi_{\TE}(\kz)\;&=\;& \arg\left[(\kzs/n_{\rm l}^2+\kzd/n_{\rm s}^2)e^{-i\kzs L}\right].\nonumber
\end{eqnarray}
The numbers $q^n_{\lambda}$ lie on the imaginary $\kz$-axis; they satisfy, cf. Eq. (\ref{eqn:ImaginaryKz}) and (\ref{trappedCond}),
\begin{equation}
\bigg(\frac{1}{n_{\rm l}^2}-1\bigg)\bkp^2<(q^n_{\lambda})^2<\bigg(\frac{1}{n_{\rm s}^2}-1\bigg)\bkp^2.
\end{equation}
The normalization constant $N_\lambda$ for trapped modes is easily obtained by direct evaluation of the integral (\ref{normTrapp}). It is given by
\begin{eqnarray}
N_\lambda=\frac{1}{2\pi}\bigg[2n_{\rm l}^2L+F_\lambda(n_{\rm l},n_{\rm s})+F_\lambda(n_{\rm l},1)\bigg]^{-1/2}\label{TrappedNorm}
\end{eqnarray}
with
\begin{eqnarray}
F_\lambda(n_{\rm l},n_{\rm s})=\frac{n_{\rm s}^2}{2}|\hat{\mathbf{e}}_{\lambda}(\bk_{\rm s}^-)|^2\frac{|t_\lambda^{\rm ls}|^2}{|\kzd|}\hspace{2 cm}
\nonumber\\
-\frac{n_{\rm l}}{\kzs}\rm{Im}(r_\lambda^{\rm ls})\;\hat{\mathbf{e}}^*_{\lambda}(\bk_{\rm l}^+)\cdot\hat{\mathbf{e}}_{\lambda}(\bk_{\rm l}^-)
\nonumber
\end{eqnarray}
and the reader is reminded that in (\ref{TrappedNorm}) the $z$-components of the wave vectors $\bk$ and $\bk_{\rm s}$ are pure imaginary and because of that the TM polarization vectors $\hat{\mathbf{e}}_{\TM}(\bk^-)$ and $\hat{\mathbf{e}}_{\TM}(\bk_{\rm s}^-)$ are no longer normalized to unity, i.e. $|\hat{\mathbf{e}}_{\TM}(\bk_{\rm s}^-)|^2\neq 1$.

\subsection{Field operators and commutation relations. Completeness of the modes.}\label{sec:Comp}
\begin{figure}[ht]
\includegraphics[width=8.5 cm, height=6.5 cm]{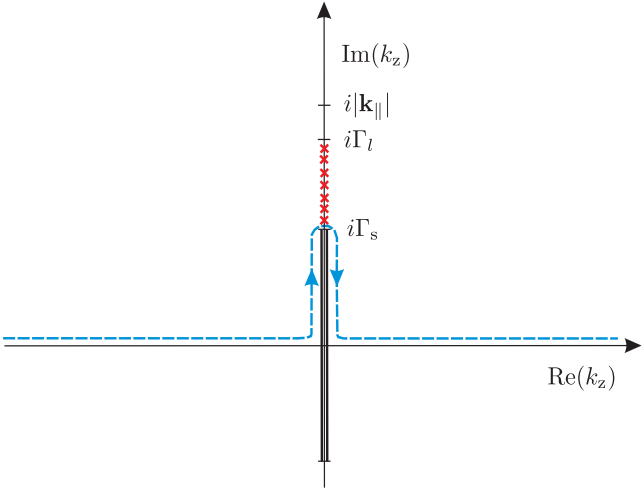}
\caption{\label{fig:gamma} The dashed line represents the contour $\gamma_{\rm s}$ used to evaluate the $\kz$ integral in Eq. (\ref{combined}). Here $\Gamma_{\rm s}=\sqrt{(n^2_{\rm s}-1)\bkp^2}/n_{\rm s}$ and $\Gamma_{\rm l}=\sqrt{(n^2_{\rm l}-1)\bkp^2}/n_{\rm l}$. The crosses represent the poles of the reflection coefficient $R_\lambda^R$ i.e. the solutions to the dispersion relation (\ref{disp}).}
\end{figure}

Now that we have determined the spatial dependence of the mode functions we are in position to write out the vector potential field operator explicitly
\begin{eqnarray}
\hat{\mathbf{A}}(\br,t)=\bigg\{ \int\rd^2\bkp 
 \int_0^\infty \rd \kz \dfrac{1}{\sqrt{2\epsilon_0\omega_\bk}}\mathbf{f}^R_{\bk\lambda}(\br)a^R_{\bk\lambda}e^{-i\omega_\bk t}\nonumber\\ 
 +\int\rd^2\bkp \int_0^\infty \rd \kzd \dfrac{1}{\sqrt{2\epsilon_0\omega_\bk}}\mathbf{f}^L_{\bk\lambda}(\br)a^L_{\bk\lambda}e^{-i\omega_\bk t}\nonumber\\
 +\int\rd^2\bkp \sum_{\kzs}\dfrac{1}{\sqrt{2\epsilon_0\omega_\bk}}\mathbf{f}^T_{\bk\lambda}(\br)a^T_{\bk\lambda}e^{-i\omega_\bk t}\bigg\}+{\rm H.C.}\label{AFieldExpl}
\end{eqnarray}
The sum in the last term runs over the allowed values of the
$z$-component of the layer's wave vector $\kzs$, i.e. the solutions of the dispersion relation (\ref{disp}). For a given type of mode, left-incident, right-incident, or trapped, photon creation and annihilation operators appearing in (\ref{AFieldExpl}) satisfy the commutation relations (\ref{Commutators}). Commutators between photon operators corresponding to different types of modes vanish as a consequence of the orthogonality of the field modes (\ref{normTrapp}), e.g.
\begin{equation}
\left[a^L_{\bk\lambda},\left(a^R_{\bk'\lambda'}\right)^\dagger\right]=0.\label{Commutators1}
\end{equation}
We would like to verify explicitly the equal-time canonical commutation
relation between field operators, say, between the electric field operator $\hat{\mathbf{E}}(\br,t)$ and the vector potential operator $\hat{\mathbf{A}}(\br,t)$
\begin{eqnarray}
\left[\hat{A}_i(\br,t),\epsilon_0\hat{E}_j(\br',t)\right]=-i\delta^\epsilon_{ij}(\br,\br'),\;\;\;z,z'>L/2\label{CanonicalComm}
\end{eqnarray}
with $\delta^\epsilon_{ij}(\br,\br')$ given by Eq. (\ref{GenDeltaLayer})
and (\ref{GenDeltaLayer1}). To evaluate (\ref{CanonicalComm}) we shall
need the electric field operator which is easily obtained from
Eq. (\ref{AFieldExpl}) using the relation
$\mathbf{E}=-\partial_t\mathbf{A}$. Plugging in the field operators into
(\ref{CanonicalComm}) and making use of commutation relations
(\ref{Commutators}) and (\ref{Commutators1}), we find that the LHS of (\ref{CanonicalComm}) is given by
\begin{eqnarray}
{\rm LHS}=i{\rm Re}\sum_\lambda\int\rd^2\bkp
\bigg[\int_0^\infty \rd \kz \;f_{\bk\lambda, i}^R(\br)f_{\bk\lambda, j}^{*R}(\br')\nonumber\\
+\int_0^\infty \rd \kzd \;f_{\bk\lambda, i}^L(\br)f_{\bk\lambda, j}^{*L}(\br')\nonumber\\
+\sum_{\kzs} \;f_{\bk\lambda, i}^T(\br)f_{\bk\lambda, j}^{*T}(\br')\bigg].\;\;\label{CommExplicit}
\end{eqnarray}
The quantity on the right-hand side is the sum over all modes, just as prescribed by equation (\ref{comp}), and therefore we expect it to be equal to the generalized transverse delta function, Eq. (\ref{GenDeltaLayer}). This shows that the statement of the completeness of the modes (\ref{comp}) is in fact equivalent to the commutation relation (\ref{CanonicalComm}), as has been noted before in \cite{Birula}. To prove that the relation 
\begin{eqnarray}
\delta_{ij}^{\epsilon}(\br,\br')=\sum_\lambda\int\rd^2\bkp\bigg[\int_0^\infty \rd \kz \;f_{\bk\lambda, i}^R(\br)f_{\bk\lambda, j}^{*R}(\br')\nonumber\\
+\int_0^\infty \rd \kzd \;f_{\bk\lambda, i}^L(\br)f_{\bk\lambda, j}^{*L}(\br')\nonumber\\
+\sum_{\kzs} \;f_{\bk\lambda, i}^T(\br)f_{\bk\lambda, j}^{*T}(\br')\bigg]
\label{CompExplicit}
\end{eqnarray}
holds for $z, z' > L/2$ we need to work out the sum over all field
modes. To start with we carry out a change of variables in (\ref{CompExplicit}): we convert the $\kzd$-integral and the $\kzs$-sum to run over the values of $\kz$. In the case of the $\kzd$-integral this is a simple change of variables according to (\ref{kzdAskz}) 
\begin{equation}
\int_0^\infty\rd\kzd=n^2_{\rm s}\int_0^\infty\rd\kz \frac{\kz}{\kzd}+n^2_{\rm s}\int_{i\Gamma_{\rm s}}^0\rd\kz \frac{\kz}{\kzd}
\end{equation}
with $\Gamma_{\rm s}=\sqrt{(n^2_{\rm s}-1)\bkp^2}/n_{\rm s}$. Here it is
seen explicitly that the contributions from the left-incident modes
split into a travelling part and an evanescent part. The values of $\kz$
included in the last integral correspond to the condition for evanescent
modes with continuous spectrum, Eq.~(\ref{evanContCond}). In the case of the sum we change the summation over $\kzs$ to run over the values of $\kz$ as defined by equation (\ref{disp2}). Plugging in the mode functions (\ref{LeftIncident}) and (\ref{RightIncident}) into equation (\ref{CompExplicit}) and utilizing straightforward properties of the reflection and transmission coefficients that hold for real $\kz, \kzd$,
\begin{equation}
R_\lambda^{*R}(-\kz)=R_\lambda^R(\kz), \;\;\;\;\frac{\kz}{\kzd}\left|T_\lambda^L\right|^2+\left|R_\lambda^R\right|^2=1,
\end{equation}
we can rewrite the completeness relation as
\begin{eqnarray}
\delta^\epsilon_{ij}(\br,\br')=\delta^\perp_{ij}(\br-\br')\hspace{5 cm}\nonumber\\
+\sum_\lambda\hat{{\rm e}}_{\lambda}^i(\mathbf{\nabla})\hat{{\rm e}}^{*j}_{\lambda}(\mathbf{\nabla'})\int\rd^2\bkp e^{i\bkp(\brp-\brp')}\hspace{2 cm}
\nonumber\\
\times\bigg\{
\sum_{q_\lambda^n}\left|N_\lambda\right|^2\left|T_\lambda^{\rm lv}\right|^2e^{i\kz(z+z')}\hspace{2.5 cm}
\nonumber\\
+\frac{1}{(2\pi)^3}\int_{i\Gamma_{\rm s}}^0\rd\kz \frac{\kz}{\kzd}\left|T_\lambda^L\right|^2 e^{i\kz(z+z')}\hspace{1 cm}\nonumber\\
+\frac{1}{(2\pi)^3}\int_{-\infty}^\infty \rd\kz R_\lambda^R e^{i\kz(z+z')}
\bigg\}.\hspace{.5 cm}\label{comp2}
\end{eqnarray}
The first term in the above equation is the standard transverse
delta-function. Therefore, if equation (\ref{CompExplicit}) is to hold,
the term in the curly brackets needs to be proportional to the
reflection part of the electrostatic Green's function, cf. the second
term on the RHS of Eq. (\ref{GenDeltaLayer1}). That this is indeed the
case is at this stage far from obvious, as for the proof one would need
to combine two integrals and a sum into one expression.
Obviously, the discreteness of the spectrum of the trapped modes is a
nuisance that needs to be overcome if one is to complete the task of
summing over the electromagnetic modes successfully. A similar
difficulty would arise in any perturbative calculation in this type of
geometry, which motivated a previous investigation of this problem for
the symmetric case of a single slab of dielectric material
\cite{completeness}. We proceed with a broadly analogous method to
\cite{completeness}, first noting that what we have here can be
considered as a superposition of a slab and a half-space geometry,
cf. \cite{completeness} and \cite {Birula}. One can utilize the
branch-cut due to $\kzd$ (which runs along the imaginary $\kz$ axis
between $\pm i\Gamma_{\rm s}$, cf.~Fig.~\ref{fig:gamma}) to express the
integral over $|T^L_\lambda|^2$ in (\ref{comp2}) as an integral over the
reflection coefficient $R_\lambda^R$ that runs from $0^-$ along the
square root cut up to the branch-point at $+i\Gamma_{\rm s}$ and then
back down to the origin $0^+$. Note that the branch-cut due to the
$\kzs$ is irrelevant because of the symmetry property of the reflection
coefficient $R_\lambda^R(-\kzs)=R_\lambda^R(\kzs)$. In this way, the first two integrals in the curly braces in equation (\ref{comp2}) can be combined together as a single integral in the complex $\kz$ plane \cite{Birula}. This is possible because the relation
\begin{eqnarray}
\frac{\kz}{\kzd}\left|T_\lambda^L\right|^2\bigg|_{\kzd,\kzs>0}\hspace{-2mm}=R_\lambda^R\bigg|_{\kzd,\kzs>0}\hspace{-2mm}-R_\lambda^R\bigg|_{\kzd,\kzs<0}
\label{AroundCut}
\end{eqnarray} 
continues to hold for coefficients (\ref{R}) with a purely imaginary
$z$-component of the vacuum wave vector, $k_{\rm z}$
(cf. \cite{Propagator}). Thus, the contributions from the travelling and
evanescent modes can be combined into a single contour integral along
the path $\gamma_{\rm s}$ depicted in Fig. \ref{fig:gamma} and the terms
appearing in the curly brackets on the RHS of Eq.~(\ref{comp2}) become
\begin{eqnarray}
\frac{1}{(2\pi)^3}\int_{\gamma_{\rm s}} \rd\kz  R_\lambda^R\hat{e}^i_\lambda(\bk^+)\hat{e}^j_\lambda(\bk^-) e^{i\kz(z+z')}\hspace{2 cm}\nonumber\\
+\sum_{q_\lambda^n}\left|N_\lambda\right|^2\left|T_\lambda^{\rm lv}\right|^2\hat{e}^i_\lambda(\bk^+)\hat{e}^j_\lambda(\bk^-)e^{i\kz(z+z')}.\;\;\;
\label{combined}
\end{eqnarray}
Here we have now included the polarization vectors explicitly in the
integrals, which is a crucial step as they affect the analytical
structure of the integrand in the complex $k_z$-plane. In particular, the TM polarization vector introduces a pole at the points $k_{\rm z}=\pm i|\bkp|$ due to the factor $1/|\bk|^2$ in its normalization factor. We will see that it is precisely this pole that gives rise to the reflection term in (\ref{GenDeltaLayer}).
We note that, according to Eq.~(\ref{R}), the reflection coefficient
contains the phase factor $e^{-i\kz L}$. Thus, since {${z+z'-L > 0}$},
the argument of the exponential in (\ref{combined}) has a negative real part in the upper half of the complex $k_z$ plane and we can evaluate the $\kz$-integral in Eq.~(\ref{combined}) by closing the contour in the upper half-plane. For this we need to determine the analytical properties of $R_\lambda^R$. We note that the denominator of the reflection coefficient (\ref{R}) is precisely the dispersion relation (\ref{disp}). Rewriting the reflection coefficients in the form
\begin{eqnarray}
R_\TE^R &=& \dfrac
{\kz-k_{\rm zl}
\left(\dfrac{1-r_\TE^{\rm ls}\exp(2ik_{\rm zl}L)}{1+r_\TE^{\rm ls}\exp(2ik_{\rm zl}L)}\right)}
{\kz+k_{\rm zl}\left(\dfrac{1-r_\TE^{\rm ls}\exp(2ik_{\rm zl}L)}{1+r_\TE^{\rm ls}\exp(2ik_{\rm zl}L)}\right)},\nonumber\\
R_\TM^R &=& \dfrac
{\kz-\dfrac{k_{\rm zl}}{n^2_{\rm l}}
\left(\dfrac{1-r_\TM^{\rm ls}\exp(2ik_{\rm zl}L)}{1+r_\TM^{\rm ls}\exp(2ik_{\rm zl}L)}\right)}
{\kz+\dfrac{k_{\rm zl}}{n^2_{\rm l}}\left(\dfrac{1-r_\TM^{\rm ls}\exp(2ik_{\rm zl}L)}{1+r_\TM^{\rm ls}\exp(2ik_{\rm zl}L)}\right)},\nonumber
\end{eqnarray}
allows us to deduce that $R_\lambda^R$ has a finite number of simple poles on the imaginary axis. When closing the contour we enclose all of them and by Cauchy's theorem the problem is reduced to the evaluation of the residues at these points: 
\begin{eqnarray}
\sum_\lambda\int_{\gamma_{\rm s}} \rd\kz R_\lambda^R \hat{e}^i_\lambda(\bk^+)\hat{e}^j_\lambda(\bk^-)e^{i\kz(z+z')}\hspace{2.7 cm}\nonumber\\
=2\pi i \sum_\lambda \sum_{\rm Res} R^R_\lambda \hat{e}^i_\lambda(\bk^+)\hat{e}^j_\lambda(\bk^-)e^{i\kz(z+z')}\hspace{2 cm}\nonumber\\
=2\pi i \bigg[\sum_\lambda\sum_{q^n_\lambda}\lim_{{\kz}\rightarrow q^n_\lambda} (\kz-q_\lambda^n)+\lim_{{\kz}\rightarrow i|\bkp|} (\kz-i|\bkp|)\bigg]\nonumber\\
\times\hat{e}^i_\lambda(\bk^+)\hat{e}^j_\lambda(\bk^-)\frac{r_\lambda^{\rm vl}+r_\lambda^{\rm ls}e^{2ik_{\rm zl}L}}{1+r_\lambda^{\rm vl}r_\lambda^{\rm ls}e^{2ik_{\rm zl}L}} e^{i\kz(z+z'-L)}\hspace{0.3 cm}
\end{eqnarray} 
Here, the first term represents the contributions from the poles in the
reflection coefficient and corresponds the trapped modes, whereas the
second term represents the contribution from the pole that arises due to
the TM polarization vector. When calculating the residues explicitly one
needs to remember that the two independent variables are $k_{\rm z}$ and
$\bkp$ and that, according to Eq. (\ref{kzsAskz}) and (\ref{kzdAskz}),
$k_{\rm zl}$ and $k_{ \rm zs}$ are functions of those. In addition, the
denominator of the reflection coefficient is not of the form
$f(\kz)(\kz-q^n_\lambda)$ so that multiplying it by $(\kz-q^n_\lambda)$
does not remove its singularity; the whole expression is still
indeterminate. Therefore, L'Hospital's rule needs to be used to evaluate
the limit (cf.\cite[Section V]{completeness}). Doing so, we find that
\begin{eqnarray}
&&\frac{1}{(2\pi)^3}\int_{\gamma_{\rm s}} \rd\kz R_\lambda^R \hat{e}^i_\lambda(\bk^+)\hat{e}^j_\lambda(\bk^-)e^{i\kz(z+z')}\nonumber\\
&&=-\sum_{q_\lambda^n}\left|N_\lambda\right|^2\left|T_\lambda^{\rm lv}\right|^2\hat{e}^i_\lambda(\bk^+)\hat{e}^j_\lambda(\bk^-)e^{i\kz(z+z')}\nonumber\\
&&\hspace*{5mm}-\nabla_i\nabla_j' G_H(\br,\br')
\label{residues}
\end{eqnarray}
where $G_H(\br,\br')$ is the reflected part of the Green's function of
the Poisson equation given in Eq.~(\ref{GenDeltaLayer1}) and derived in
Appendix \ref{App:Elstatic}. We see that the poles of the reflection coefficient $R_\lambda^R$ yield a term that exactly cancels out the contributions of the trapped modes to the completeness relation (\ref{comp2}) whereas the pole of the $\TM$ polarization vector yields the term proportional to Green's function. Thus, the final result can be written as
\begin{eqnarray}
&&\int \rd^2\bkp \sum_{k_z} \hspace{-6mm}\;\int f^i_{\bk\lambda}(\br)f^{*j}_{\bk\lambda}(\br')=\frac{1}{i}[A_i(\br),-\epsilon_0E_j(\br')]\nonumber\\
&&=\delta^\perp_{ij}(\br-\br')-\nabla_i\nabla_j' G_H(\br,\br')\;\;\;z,z'>L/2\nonumber\\
&&=\delta_{ij}\delta^{(3)}(\br-\br')-\nabla_i\nabla_j'
G(\br,\br')\;\;\;z,z'>L/2\nonumber
\end{eqnarray}
which is precisely what we have anticipated earlier. In the next section we demonstrate how the calculation presented here may be applied to accomplish typical perturbative QED calculations in a layered geometry.

\section{Energy shift}\label{sec:Shift}
\begin{figure}[ht]
\includegraphics[width=8.5 cm, height=6.7 cm]{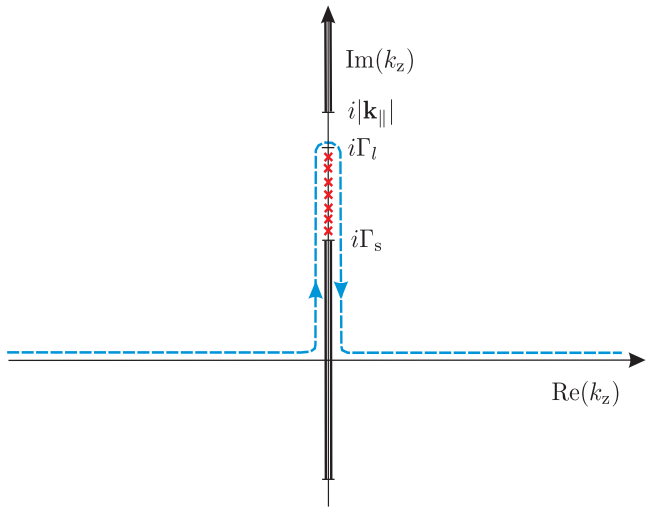}
\caption{\label{fig:d}The dashed line represents the final contour $\gamma_{\rm l}$ used to evaluate the energy shift in Eq. (\ref{shift2}).}
\end{figure}

To work out the energy shift we use standard perturbation theory where
the atom is treated by means of the Schr${\rm \ddot{o}}$dinger quantum
mechanics and only the electromagnetic field is second-quantized. We
work with a multipolar coupling where the lowest order of the interaction Hamiltonian is
\begin{equation}
H_{\rm int}=-\boldsymbol{\mu}\cdot\mathbf{E} .
\end{equation}
Then the energy shift of the atomic state $i$, up to the second-order, is given by
\begin{equation}
\Delta E_i=\langle i;0|H_{\rm int}| i; 0\rangle+\sum_{j\neq i}\sum_{\bk, \lambda} \hspace{-5mm}\int\;\;\frac{|\langle j;0|H_{\rm int}| i; 1_{\bk\lambda}\rangle|^2}{E_i-(E_j+\omega_\bk)}.\nonumber
\end{equation}
Here, $\boldsymbol{\mu}$ is the atomic electric dipole moment, and the
composite state $|j;1_{\bk\lambda}\rangle$ describes the atom in the
state $|j\rangle$ with energy $E_j$ and the photon field containing one
photon with momentum $\bk$ and polarization $\lambda$. Because the
electric field operator is linear in the photon creation and
annihilation operators, the first-order contribution vanishes and the
second-order correction is the lowest-order contribution. Since the
electric field does not vary appreciably over the size of the atom we
use the electric dipole approximation. Then the energy shift can be expressed as
\begin{equation}
\Delta E_i=-\sum_{j\neq i}\sum_{\bk, \lambda} \hspace{-5mm}\int\;\;\frac{\omega_\bk}{2\epsilon_0}\frac{|\langle i|\boldsymbol{\mu}|j\rangle\cdot\mathbf{f}^*_{\bk\lambda}(\br_0) |^2}{E_{ji}+\omega_\bk}\label{Pert}
\end{equation}
where $\br_0=(0,0,z_0)$ is the position of the atom and we have abbreviated $E_{ji}=E_j-E_i$. It is seen that the calculation involves a summation over the modes of the electromagnetic field as carried out in the proof of the completeness relation (\ref{comp2}). Equation (\ref{Pert}) can be written out explicitly as
\begin{eqnarray}
\Delta E_i=-\frac{1}{2\epsilon_0}\sum_{\lambda}\sum_{j\neq i}|\mu_m|^2\int\rd\bkp\hspace{1.8 cm}\nonumber\\\times
\left(\Delta^{\rm vac}+\Delta^{\rm trav}+\Delta^{\rm evan}+\Delta^{\rm trap}\right)\hspace{.2 cm}\label{shift1}
\end{eqnarray}
with $|\mu_m|^2\equiv|\langle i|\mu_m|j\rangle|^2$. There are four
distinct contributions to the energy shift. $\Delta^{\rm vac}$ is the
position-independent contribution caused by the vacuum fields and gives rise to the Lamb shift in free space
\begin{eqnarray}
\Delta^{\rm vac}=\frac{1}{(2\pi)^3}\int_{-\infty}^\infty\rd\kz\;
e_\lambda^m(\bk^-)e_\lambda^{m*}(\bk^-)\frac{\omega}{E_{ji}+\omega} . 
\label{vacuum}
\end{eqnarray}
The remaining three contributions come from the travelling, evanescent, and trapped modes, respectively,
\begin{eqnarray}
\Delta^{\rm trav}\hspace{-2mm}&=&\hspace{-2mm}\frac{1}{(2\pi)^3}\int_{-\infty}^\infty\rd\kz R_\lambda^Re_\lambda^m(\bk^+)e_\lambda^{m*}(\bk^-)e^{2i\kz z_0}
\frac{\omega}{E_{ji}+\omega},\nonumber\\
\Delta^{\rm evan}\hspace{-2mm}&=&\hspace{-2mm}\frac{1}{(2\pi)^3}
 \int_{i\Gamma_{\rm s}}^0\rd\kz\;
 \frac{\kz}{\kzd}|T_\lambda^L|^2e_\lambda^m(\bk^+)e_\lambda^{m*}(\bk^+)e^{2i\kz
 z_0}\nonumber\\
&&\hspace*{5mm}\times\frac{\omega}{E_{ji}+\omega},\label{remaining}\\
\Delta^{\rm trap}\hspace{-2mm}&=&\hspace{-2mm}\sum_{q_\lambda^n}|N_\lambda|^2|T_\lambda^{\rm lv}|^2e_\lambda^m(\bk^+)e_\lambda^{m*}(\bk^+)e^{2i\kz z_0}\frac{\omega}{E_{ji}+\omega},\nonumber
\end{eqnarray}
with $z_0$ being the position of the atom with respect to the origin. Note that because of the dipole approximation the shorthand notation for polarisation vectors (\ref{TM}) can be no longer applied. Normally one is interested in the energy shift caused by the presence of the dielectric boundaries only i.e. the correction to the shift that would appear in the free space. Therefore, we renormalize the energy-level shift (\ref{shift1}) by subtracting from it its free space limit, i.e.
\begin{equation}
\Delta E_i^{\rm ren}=\Delta E_i - \lim_{n_{\rm l},n_{\rm s}\rightarrow 1} \Delta E_i\;.
\end{equation}
The renormalization procedure amounts to the removal of the
contributions $\Delta^{\rm vac}$, Eq.~(\ref{vacuum}), from the energy
shift (\ref{shift1}) and takes care of any infinities that would appear
otherwise, provided we treat the remaining parts with care. As noted
elsewhere \cite{slab}, the contributions (\ref{remaining}) suffer from
convergence problems when treated separately. However, appropriate tools
to handle the problem have been developed in Sec. \ref{sec:Comp}. We aim
to combine $\Delta^{\rm trav}$, $\Delta^{\rm evan}$ and $\Delta^{\rm trap}$ 
into one compact expression that is easy to handle analytically. We can
use the same trick as in the proof of the completeness relation because
the analytical structure of the integrand in the complex $\kz$-plane is
the same except for the function $\omega=(\bkp^2+\kz^2)^{1/2}$ that
comes about due to the denominator of perturbation theory and introduces
additional branch-points at $\kz=\pm i|\bkp|$ as compared to
Fig. \ref{fig:gamma}. This poses no difficulties though, if one chooses
the branch-cuts to lie between $\pm i|\bkp|$ and $\pm i\infty$. Then,
the contributions to the energy shift from the travelling modes
$\Delta^{\rm trav}$ and the evanescent modes $\Delta^{\rm evan}$ can be
combined together into a single complex integral as explained in the
steps between Eq. (\ref{comp2}) and Eq. (\ref{combined}). This is
possible because for imaginary $\kz$ we have
$e_\lambda^{m*}(\bk^+)=e_\lambda^{m}(\bk^-)$, whereas for real $\kz$ the
relation $e_\lambda^{m*}(\bk^-)=e_\lambda^{m}(\bk^-)$ holds. On the other hand, we also know from Eq. (\ref{residues}) that the sum in $\Delta^{\rm trap}$ is equal to an integral over the reflection coefficient $R_\lambda^R$ taken along any clockwise contour enclosing all of it's poles. Choosing this contour to run from $\kz=0^-+i\Gamma_{\rm s}$ to $\kz=0^-+i\Gamma_{\rm l}$ and then back down from $\kz=0^++i\Gamma_{\rm l}$ to $\kz=0^++i\Gamma_{\rm s}$, cf. Fig. \ref{fig:d}, we write down the renormalized energy shift compactly as
\begin{eqnarray}
&&\Delta E_i^{\rm ren}=-\frac{1}{2(2\pi)^3\epsilon_0}\sum_{m,\lambda}\sum_{j\neq i}|\mu_m|^2\int\rd\bkp\nonumber\\
&&\hspace{5mm}\times\int_{\gamma_{\rm l}}\rd\kz\;\frac{\omega}{E_{ji}+\omega}\;R_\lambda^Re_\lambda^m(\bk^+)e_\lambda^{m}(\bk^-)e^{2i\kz z_0} \label{shift2}
\end{eqnarray}
where the contour of integration $\gamma_{\rm l}$ is shown in {Fig. \ref{fig:d}}. It resembles that of {Fig. \ref{fig:gamma}} but now runs on the imaginary axis up to the point $\kz=i\Gamma_{\rm l}$ enclosing all the poles of the reflection coefficients $R^R_\lambda$. 
Formula (\ref{shift2}) is equally applicable to ground-state atoms
$| 0\rangle$ as it is to atoms that are in an excited state $|i\rangle$
provided we use the contour of integration as given in
{Fig. \ref{fig:d}} and interpret the $k_z$ integral as a Cauchy
principal-value. As renormalization has now been dealt with we shall
from now on omit the superscript ``ren'' and designate the renormalized
energy shift of Eq.~(\ref{shift2}) simply by $\Delta E_i$.

\subsection{Ground state atoms}
\begin{figure}[ht]
\includegraphics[width=8.5 cm, height=5.7 cm]{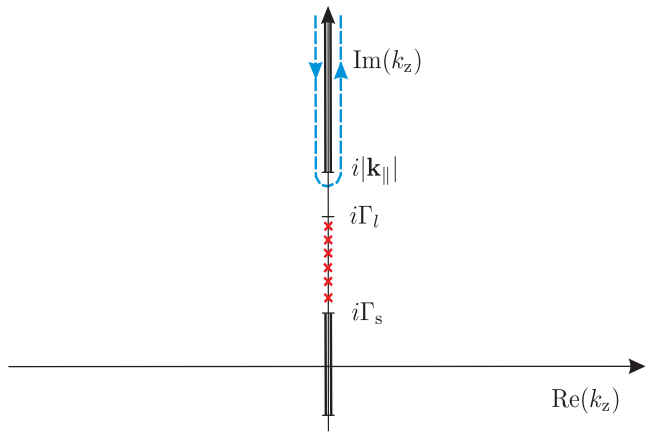}
\caption{\label{fig:c} The final contour $\mathcal{C}$ used to evaluate the energy shift of the ground state atom in Eq. (\ref{shift3}).}
\end{figure}

In the case of a ground-state atom the energy difference $E_{j0}\equiv E_j-E_0$ is always positive hence the denominator in Eq. (\ref{shift2}) that originates from second-order perturbation theory, $E_{j0}+\omega$, never vanishes. Then, Eq. (\ref{shift2}) contains no poles in the upper half of the $k_z$-plane other than those due to the reflection coefficient $R^R_\lambda$. To evaluate the $k_z$ integral we can deform the contour of integration in Eq. (\ref{shift2}) from that sketched in Fig. \ref{fig:d} to the one as shown in Fig. \ref{fig:c} which is beneficial from the computational point of view as it simplifies the analysis of Eq. (\ref{shift2}) considerably.
Writing out explicitly the sums over the polarization vectors (\ref{TM}) and then expressing the integral in the $\bkp$-plane in polar coordinates, $k_{\rm x}=k_\parallel\cos\phi,\; k_{\rm y}=k_\parallel\sin\phi$, where the angle integral is computable analytically, we rewrite the energy shift as
\begin{eqnarray}
\Delta E_0=\frac{1}{16\pi^2\epsilon_0}\sum_{j\neq 0}\int_0^\infty \rd k_\parallel\;k_\parallel \int_\mathcal{C} \rd \kz \frac{\omega }{E_{j0}+\omega}e^{2i\kz \z}\nonumber\\
\times 
\left[
|\mu_\parallel|^2\left(\tilde{R}_{\TE}^R-\frac{\kz^2}{\omega^2}\tilde{R}_{\TM}^R \right)\right.
\left.+2|\mu_\perp|^2 \frac{k_\parallel^2}{\omega^2}\tilde{R}_{\TM}^R 
\right]\label{shift3}
\end{eqnarray}
with $\omega(\kz)=\sqrt{k_\parallel^2+\kz^2}$,
$|\mu_\parallel|^2=|\mu_x|^2+|\mu_y|^2$ and the contour $\mathcal{C}$ is
that in Fig. \ref{fig:c}. The amended reflection coefficients $\tilde{R}^R_\lambda$  are given by
\begin{equation}
\tilde{R}_\lambda^R=\frac{r_\lambda^{\rm vl}+r_\lambda^{\rm ls}e^{2ik_{\rm zl}L}}{1+r_\lambda^{\rm vl}r_\lambda^{\rm ls}e^{2ik_{\rm zl}L}},\label{TildedR}
\end{equation}
i.e. we have pulled out the phase factor $e^{-i\kz L}$ in order to
define $\z=z_0-L/2$ as the distance between the atom and the surface,
cf. Eq. (\ref{R}).

In order to perform the $\kz$ integration in (\ref{shift3}) we need to analytically continue the function $\omega=\omega(\kz)$, which is real and positive on the real axis, to the both sides of the branch cut along which the integration is carried out, (cf. Fig. \ref{fig:c}). Doing so we find that on the LHS of the cut the positive value of the square root needs to be taken, and hence on the RHS of the cut we must take the opposite sign. Therefore we have
\begin{eqnarray}
\int_C \rd \kz\frac{\omega}{E_{j0}+\omega}=-\int_{iq}^{i\infty}\rd\kz\;\frac{2E_{j0}\omega}{(E_{j0}-\omega)(E_{j0}+\omega)}\;.\nonumber
\end{eqnarray}
Now we carry out a sequence of changes of variables. First we re-express
the $k_z$ integration in terms of one over the frequency $\omega$ by substituting $\omega=\sqrt{k_\parallel^2+\kz^2}$,
\begin{equation}
\int_{ik_\parallel}^{i\infty}\rd\kz=\int_0^{i\infty}\rd\omega\frac{\omega}{\sqrt{\omega^2-k_\parallel^2}}.
\end{equation}
Then, we make the integral run along the real axis by setting $\omega=i\xi$. After this is done, the energy shift of the ground state is expressed as a double integral that covers the first quadrant of the $(k_\parallel,\xi)$-plane
\begin{eqnarray}
\Delta E_0=-\frac{1}{8\pi^2\epsilon_0}\sum_{j\neq i}E_{j0}\int_0^\infty\rd k_\parallel\,k_\parallel\int_0^\infty \rd \xi \;\frac{e^{-2\sqrt{\xi^2+k_\parallel^2}\z }}{\sqrt{\xi^2+k_\parallel^2}(E_{j0}^2+\xi^2)}\nonumber\\
\times
\left\{ 
|\mu_\parallel|^2\left[(\xi^2+k_\parallel^2)\tilde{R}_{\TM}^R-\xi^2\tilde{R}_{\TE}^R \right]+2k_\parallel^2\tilde{R}_{\TM}^R |\mu_\perp|^2
\right\}.\nonumber\label{NonRetAsym}
\end{eqnarray}
It seems natural to introduce polar coordinates, $k_\parallel=\bar{x}\sin\phi,\; \xi=\bar{x}\cos\phi$. We also choose to scale the radial integration variable $\bar{x}=E_{j0}x$ with $E_{j0}>0$ and set $y=\cos\phi$. This provides us with the final form of the energy shift that is more suitable for numerical computations and asymptotic analysis
\begin{eqnarray}
\Delta E_0=\frac{1}{8\pi^2\epsilon_0}\sum_{j\neq i}E_{j0}^3\int_0^\infty\rd x x^3\int_0^1\rd y \;\frac{e^{-2E_{j0}\z x}}{1+x^2y^2}\hspace{0.5 cm}\nonumber\\
\times
\left[ 
|\mu_\parallel|^2\left(y^2\tilde{R}_{\TE}^R-\tilde{R}_{\TM}^R \right)+2|\mu_\perp|^2(y^2-1)\tilde{R}_{\TM}^R 
\right].\label{RetAsym}
\end{eqnarray}
The reflection coefficients $\tilde{R}_\lambda^R$ are as expressed in
(\ref{TildedR}) but with the wave vectors given by
\begin{eqnarray}
k_{\rm zi} = ixE_{j0}\sqrt{(n_{\rm i}^2-1)y^2+1},\;\;\;n_{\rm i}=\{1,n_{\rm l},n_{\rm s}\}.\nonumber
\end{eqnarray}
Note that even though the wave vector is imaginary, the final result is
a real number, as it should, because the Fresnel coefficients contain
only ratios of wave vectors.

\subsection{Excited atoms}
As mentioned previously, the energy-level shift of an excited atom is
also given by Eq. (\ref{shift2}). However, one needs to take account of
the fact that the quantity $E_{ji}\equiv E_j-E_i$ can now become
negative for $E_j<E_i$, so that the denominator originating from
perturbation theory contributes additional poles lying on the path of
$k_z$ integration, shown in {Fig. \ref{fig:d}} and is now to be
understood as a Cauchy principal-value. These poles are located at
$k_z=\pm\sqrt{E_{ji}^2-\bkp^2}$, though their precise location depends on
the value of $|\bkp|$ that is not fixed but varies as we carry out the
$\bkp$ integrations in equation (\ref{shift2}). For $|\bkp| \in
[0,|E_{ji}|]$ the poles are located on the real $k_z$ axis but as we
increase the value of $|\bkp|$ to exceed $|E_{ji}|$ both poles move onto
the positive imaginary axis according to the convention that
$\rm{Im}(k_z)>0$. For $|\bkp|$ belonging to the interval
$[|E_{ji}|,n_{\rm s}|E_{ji}|]$ the poles are located on the opposite
sides of the branch-cut due to the $k_{\rm zs}$ and care needs to be
taken when evaluating those pole contributions. To evaluate the Cauchy
principal-value of the $k_z$-integral we circumvent the poles and close
the contour in the upper half-plane, as was done in the previous
section. The contribution from the large semicircle vanishes and
equation (\ref{shift2}) acquires pole contributions that are easily
worked out by the residue theorem. The energy shift splits into the a
"non-resonant" ground-state-like part $\Delta E_i$ and a "resonant"
oscillatory part $\Delta E^{\rm res}_i$ that arises only if the atom is
in an excited state. In analogy to the result of the previous section, the "non-resonant" part is given by
\begin{eqnarray}
\Delta E_i=\frac{1}{8\pi^2\epsilon_0}\sum_{j\neq i}E_{ji}^3\int_0^\infty\rd x x^3\int_0^1\rd y \;\frac{e^{-2|E_{ji}|\z x}}{1+x^2y^2}\hspace{0.5 cm}\nonumber\\
\times
\left[ 
|\mu_\parallel|^2\left(y^2\tilde{R}_{\TE}^R-\tilde{R}_{\TM}^R \right)+2|\mu_\perp|^2(y^2-1)\tilde{R}_{\TM}^R 
\right]\label{Excited1}
\end{eqnarray}
with wave vectors expressed as
\begin{equation}
k_{\rm zi}= ix|E_{ji}|\sqrt{(n_{\rm i}^2-1)y^2+1},\;\;\;n_{\rm i}=\{1,n_{\rm l},n_{\rm s}\},
\end{equation}
whereas the "resonant" part is given by
\begin{eqnarray}
\Delta E^{\rm res}_i={\rm Re}\frac{i}{8\pi\epsilon_0}\sum_{j
 <i}|E_{ji}|^3\int_0^\infty\hspace{-1 mm}\frac{\rd q\, q}{\sqrt{1-q^2}}e^{2i|E_{ji}|\sqrt{1-q^2}\z}\hspace{0 cm}\nonumber\\
\times \left\{
|\mu_\parallel|^2\left[(1-q^2)\tilde{R}_{\TM}^{R}-\tilde{R}_{\TE}^{R}\right]-2|\mu_\perp|^2q^2\tilde{R}_{\TM}^{R} 
\right\},\;\;\;\label{Excited2}
\end{eqnarray}
with wave vectors expressed as
\begin{eqnarray}
k_{\rm zi}= |E_{ji}|\sqrt{n_{\rm i}^2-q^2},\;\;\;n_{\rm i}=\{1,n_{\rm l},n_{\rm s}\}.\nonumber
\end{eqnarray}
The reflection coefficients are as given in (\ref{TildedR}). The integral in Eq.~(\ref{Excited2}) contains poles because the dispersion relation present in the denominators of the reflection coefficients has now solutions on the real axis when $q\in[n_{\rm s}, n_{\rm l}]$. This signals contributions from surface excitations (trapped modes). This fact has been mentioned in \cite{sipe} where the interaction of an excited atom with layered dielectric has been studied, although using mainly numerical analysis. Here we will attempt to study the results (\ref{Excited1}) and (\ref{Excited2}) analytically. To do so it will prove beneficial to rewrite equation (\ref{Excited2}) slightly. We change variables according to $\sqrt{1-q^2}=\eta$ and split the contributions to Eq. (\ref{Excited2}) into two parts. The first one is a contribution from the travelling modes and given by
\begin{eqnarray}
\Delta E^{\rm res, trav}_i=-{\rm Re}\frac{i}{8\pi\epsilon_0}\sum_{j <i}|E_{ji}|^3\int_0^1\rd \eta e^{2i|E_{ji}|\z \eta}\hspace{0.7 cm}\nonumber\\
\times \left\{
|\mu_\parallel|^2\left[\tilde{R}_{\TE}^{R}-\eta^2\tilde{R}_{\TM}^{R} \right]+2|\mu_\perp|^2(1-\eta^2)\tilde{R}_{\TM}^{R} 
\right\}\;\;\;\label{ExcitedElstatic1}
\end{eqnarray}
where the wave vectors in reflection coefficients are all real and can be expressed as 
\begin{equation}
k_{\rm zi}=|E_{ji}|\sqrt{n_{\rm i}^2-1+\eta^2},\;\;\;n_{\rm i}=\{1,n_{\rm l},n_{\rm s}\},
\end{equation}
and the second is a contribution from the evanescent modes 
\begin{eqnarray}
\Delta E^{\rm res, evan}_i=-{\rm Re}\frac{1}{8\pi\epsilon_0}\sum_{j <i}|E_{ji}|^3\int_0^\infty\rd \eta e^{-2|E_{ji}|\z \eta}\hspace{0.5 cm}\nonumber\\
\times \left\{
|\mu_\parallel|^2\left[\tilde{R}_{\TE}^{R}+\eta^2\tilde{R}_{\TM}^{R} \right]+2|\mu_\perp|^2(1+\eta^2)\tilde{R}_{\TM}^{R} 
\right\}\;\;\;\label{ExcitedElstatic2}
\end{eqnarray}
where the wave vectors in reflection coefficients can be expressed as 
\begin{equation}
k_{\rm zi}=|E_{ji}|\sqrt{n_{\rm i}^2-1-\eta^2},\;\;\;n_{\rm i}=\{1,n_{\rm l},n_{\rm s}\}.\label{keta}
\end{equation}
Finally, it is worth noting that the imaginary part of
Eq. (\ref{Excited2}) is actually proportional to the modified decay
rates \cite{Wu}. These have already been studied in \cite{Urbach} so
that we focus on energy shifts only. However, the methods of analysis
that are reported in the next section do allow one to write down at once
equivalent analytical formulae for the decay rates.

\section{Asymptotic analysis}\label{sec:Asym}
The interaction between the atom and the dielectric is electromagnetic
in nature and it is mediated by photons. The atomic system in state
$|i\rangle$ evolves in time with a characteristic time-scale that is
proportional to $E_{ji}^{-1}$, with $E_{ji}$ being the energy-level
spacing between the states $|i\rangle$ and $|j\rangle$ which are
connected by the strongest dipole transition from state
$|i\rangle$. Since it takes a finite time for the photon to make a round
trip between the atom and the surface, the atom will have changed by the
time the photon comes back. Therefore, the ratio of the time needed by
the photon to travel to the surface and back and the typical time-scale
of atomic evolution is a fundamental quantity that plays decisive role
in characterising the interaction. In natural units, if $2E_{ji}\z \ll 1$
we can safely assume that the interaction is instantaneous and we are in
the so-called non-retarded or van der Waals regime. If $2E_{ji}\z \gg 1$
the interaction becomes manifestly retarded as the atom will have
changed significantly by the time the photon comes back. However, the
problem we have considered here provides us with yet another length
scale, namely the thickness of the top layer $L$. We shall now consider
the energy shift in various asymptotic regimes.

\subsection{Ground state atoms. Electrostatic limit, ($2E_{ji}\z\ll 1$)}
In this limit the interaction is instantaneous (or electrostatic) in
nature and the energy shift is obtainable using the Green's function of
the classical Laplace equation (cf. e.g. \cite{nonret}). This classical
derivation is outlined in the Appendix \ref{App:Elstatic}. The end
result for the energy shift reads
\begin{eqnarray}
\Delta E^{\rm el}=-\frac{1}{16\pi\epsilon_0}\left(\langle\mu_\parallel^2\rangle+2\langle\mu_\perp^2\rangle\right)  \int_0^\infty\rd k k^2e^{-2k\z}\hspace{1 cm}\nonumber\\
\times\left(\dfrac{\dfrac{n_{\rm l}^2-1}{n_{\rm l}^2+1}-\dfrac{n_{\rm l}^2-n_{\rm s}^2}{n_{\rm s}^2+n_{\rm l}^2}e^{-2kL}}{1-\dfrac{n_{\rm l}^2-1}{n_{\rm l}^2+1}\dfrac{n_{\rm l}^2-n_{\rm s}^2}{n_{\rm s}^2+n_{\rm l}^2}e^{-2kL}}\right),\hspace{2 mm}\label{elstatic}
\end{eqnarray}
with $\langle
\mu_\parallel^2\rangle\equiv\langle\mu_x^2\rangle+\langle\mu_y^2\rangle$
and $\langle\mu_\perp^2\rangle\equiv\langle\mu_z^2\rangle$. We will now
show that one can also obtain the above result as a limiting case of the
results of previous section, thus providing a cross-check for our
general calculation. To start with we note that equation (\ref{RetAsym})
cannot be used to take the electrostatic limit in which we
mathematically let $E_{ji}\rightarrow 0$ because it has been scaled with
$E_{ji}$. Therefore, it is best to start from equation
(\ref{shift2}). The result of Eq.~(\ref{elstatic}) can be derived very
quickly if we observe that in the limit $E_{ji}\rightarrow 0$ the branch
cut due to $\omega=\sqrt{\bkp^2+k_{\rm z}^2}$ is no longer present and
the contour in Fig. \ref{fig:c} collapses to a simple enclosure of the
point $\kz=i|\bkp|$. The contribution from the TE mode vanishes as the
product of the polarization vectors is regular at $\kz=i|\bkp|$, but for
the TM mode this point is a simple pole, cf. Eq. (\ref{TM}). Therefore
we obtain
\begin{eqnarray}
\Delta E^{\rm el}=-\frac{1}{(2\pi)^32\epsilon_0}\sum_m \sum_{j\neq i} |\mu_m|^2 
\int\rd\bkp  \hspace{2.5 cm}\nonumber\\
\times  2\pi i\lim_{\kz\rightarrow i|\bkp|}(\kz-i|\bkp|)R_{\TM}^R e_{\TM}^m(\bk^+)e_{\TM}^m(\bk^-)e^{2i\kz z_0}. \nonumber
\end{eqnarray}
Taking the limit and expressing the remaining integrals in polar coordinates, where the angle integral is elementary, yields equation (\ref{elstatic}) with $\langle\mu_m^2\rangle\equiv\sum_{j\neq i} |\langle i|\mu_m|j\rangle|^2=\langle i|\mu_m^2|i\rangle$. Equation (\ref{elstatic}) can be further analysed depending on the relative values of $L$ and $\mathcal{Z}$.

\subsubsection{Thin layer ($\z/L\gg 1$)}
In this case the distance of the atom from the surface is much greater than the thickness of the layer of refractive index $n_{\rm l}$ (but still small enough for the retardation to be neglected). Then, rescaling the integral in equation (\ref{elstatic}) with $k=x/L$ allows us to use  Watson's lemma \footnote{The essential idea is to spot that, since the integrand is strongly damped by the exponential, most of the contributions to the integral will come from small values of $k$. Thus, it is permissible to Taylor-expand the remaining part of the integrand about $k=0$. For a more rigorous treatment see \cite{Bender}.} to derive the following result
\begin{eqnarray}
\Delta E^{\rm el}\approx\Delta E^{\rm el}_{n_{\rm s}}-\frac{1}{64\pi\epsilon_0\z^3}\left(\langle\mu_\parallel^2\rangle+2\langle\mu_\perp^2\rangle\right)\hspace{1 cm}
\nonumber\\
\times\left[a_1\frac{L}{\z}+a_2\frac{L^2}{\z^2}+O\left(\frac{L^3}{\z^3}\right)\right],\label{ElstaticLargeD}
\end{eqnarray}
with the coefficients $a_i$ given by
\begin{eqnarray}
a_1\;&=\;&\frac{3}{n_{\rm l}^2}\frac{n_{\rm l}^4-n_{\rm s}^4}{(n_{\rm s}^2+1)^2}\;,\nonumber\\
a_2\;&=\;&-\frac{6}{n_{\rm l}^4}\frac{(n_{\rm l}^4-n_{\rm s}^4)(n_{\rm s}^2+n_{\rm l}^4)}{(n_{\rm s}^2+1)^3}\;,\nonumber
\end{eqnarray}
where $\Delta E^{\rm el}_{n_{\rm s}}$ is the well-known electrostatic interaction energy between an atom and a dielectric half-space of refractive index $n_{\rm s}$ that can be obtained by the method of images 
\begin{equation}
\Delta E^{\rm el}_{n_{\rm s}}=-\frac{1}{64\pi\epsilon_0 \z^3}\frac{n_{\rm s}^2-1}{n_{\rm s}^2+1}\left(\langle\mu_\parallel^2\rangle+2\langle\mu_\perp^2\rangle\right).\label{PlaneElstat}
\end{equation}
The corrections to this result are represented by the remaining elements of the asymptotic series. Note that if $n_{\rm l}>n_{\rm s}$ then $a_1>0$ and, not surprisingly, the interaction, as compared to a half-space alone, is enhanced by the presence of the thin dielectric layer of higher refractive index $n_{\rm l}$.

\subsubsection{Thick layer ($\z/L\ll 1$)}\label{Sect:El:smalld}
In this case the thickness of the layer is much greater than the distance between the atom and the surface.  The top layer now appears from the point of view of the atom almost as a half-space of refractive index $n_{\rm l}$ only that it is in fact of finite thickness. To analyse the result (\ref{elstatic}) in this limit we cast it in a somewhat different form. Note that, especially when $kL$ is large but not only then,
\begin{equation}
\frac{n_{\rm l}^2-1}{n_{\rm l}^2+1}\frac{n_{\rm l}^2-n_{\rm s}^2}{n_{\rm s}^2+n_{\rm l}^2}e^{-2kL}<1
\end{equation}
and the denominator of the integrand in Eq. (\ref{elstatic}) can be written as geometrical series. Since the series is absolutely convergent we can integrate it term by term and obtain the following representation of the electrostatic result
\begin{eqnarray}
\Delta E^{\rm el}=\Delta E^{\rm el}_{n_{\rm l}}+\frac{1}{16\pi\epsilon_0}
\left(\langle\mu_\parallel^2\rangle+2\langle\mu_\perp^2\rangle\right) 
\dfrac{n_{\rm l}^2}{n_{\rm l}^4-1}\hspace{1.0 cm}
\nonumber\\
\times\sum_{\nu=1}^\infty\left(\dfrac{n_{\rm l}^2-1}{n_{\rm l}^2+1}\dfrac{n_{\rm l}^2-n_{\rm s}^2}{n_{\rm s}^2+n_{\rm l}^2}\right)^\nu\frac{1}{(\z+\nu L)^3}\;\;\;\;\label{sumsol}
\end{eqnarray}
where $\Delta E^{\rm el}_{n_{\rm l}}$ is the electrostatic energy shift due to a single half-space of refractive index $n_{\rm l}$, i.e. Eq. (\ref{PlaneElstat}) with $n_{\rm s}$ replaced by $n_{\rm l}$. The sum in Eq. (\ref{sumsol}) represents the correction to $\Delta E^{\rm el}_{n_{\rm l}}$ due to the finite thickness of the layer. For fixed $\z$ and $L$ it can be easily computed numerically to any desired degree of accuracy. We note however, that to the leading order in $\z /L$ the interaction is weakened by the same amount independently of the distance of the atom from the surface and therefore is not measurable. The next-to-leading order correction is the first to be distance-dependent and is proportional to $\z /L^4$, which can be easily seen by expanding the factor in series around $\z/\nu L = 0$:
\begin{equation}
\frac{1}{(\z+\nu L)^3}\approx\frac{1}{\nu^3L^3}-\frac{3Z}{\nu^4L^4}+O\left(\frac{Z^2}{L^5}\right).
\end{equation}

\subsection{Ground state atoms. Retarded limit, ($2\z E_{ji}\gg 1$)}
\subsubsection{Thin layer ($\z/L \gg  1$)}
In this case we study the situation when the top layer is much thinner than the distance between the atom and the surface. To obtain the asymptotic series we use Watson's lemma in much the same way as in the electrostatic case \cite{Bender}. Series expansion of the integrand in Eq. (\ref{RetAsym}) about $x=0$ decouples the integrals and the resulting integrals can be calculated analytically. Thus, to first approximation, for an atom located sufficiently far from the interface, the impact of the thin dielectric layer on the standard Casimir-Polder interaction can be described by 
\begin{eqnarray}
\Delta E^{\rm ret}=\Delta E_{n_{\rm s}}^{\rm ret}\hspace{6.0 cm}\nonumber\\
-\frac{1}{16\pi^2\epsilon_0\z^4}\sum_{j\neq i}
\left[\frac{a_\parallel|\mu_\parallel|^2+2a_\perp|\mu_\perp|^2}{E_{ji}}\right]\frac{L}{\z}
+O\left(\frac{L^2}{\z^2}\right)\;\;\;\label{LargeDAsym}
\end{eqnarray} 
where $\Delta E_{n_{\rm s}}^{\rm ret}$ is the retarded limit of energy shift as caused by a single dielectric half-space of refractive index $n_{\rm s}$, which was calculated in \cite{Wu}. We give this result in Appendix \ref{App:HSResult}. The coefficients $a_\parallel$ and $a_\perp$ in (\ref{LargeDAsym}) can be expressed in terms of elementary functions as
\begin{widetext}
\begin{eqnarray}
a_\parallel\;&=\;&\frac{1}{n_{\rm l}^2}\frac{n_{\rm l}^2-n_{\rm s}^2}{(n_{\rm s}^2-1)^2(n_{\rm s}^2+1)}\bigg[n_{\rm s}^5(6n_{\rm s}-3)(n_{\rm l}^2-1)+3n_{\rm s}^2(n_{\rm l}^2+1)-n_{\rm l}^2(2n_{\rm s}^4+3n_{\rm s}^3+3n_{\rm s}-8)\bigg]\nonumber\\
\;&-\;&\frac{n_{\rm l}^2-n_{\rm s}^2}{n_{\rm l}^2(n_{\rm s}^2-1)^{5/2}}\ln\left(\sqrt{n_{\rm s}^2-1}+n_{\rm s}\right)
\bigg[ 2n_{\rm s}^2n_{\rm l}^2(n_{\rm s}^2-1)^2-2n_{\rm s}^4(n_{\rm s}^2-1)+n_{\rm l}^2 \bigg]\nonumber\\
\;&-\;&\frac{n_{\rm s}^4}{2n_{\rm l}^2}\frac{n_{\rm l}^2-n_{\rm s}^2}{(n_{\rm s}^2-1)^2(n_{\rm s}^2+1)^{3/2}}
\ln\left(\frac{\sqrt{n_{\rm s}^2+1}+1}{\sqrt{n_{\rm s}^2+1}-1}\frac{\sqrt{n_{\rm s}^2+1}-n_{\rm s}}{\sqrt{n_{\rm s}^2+1}+n_{\rm s}}\right)
\bigg[2n_{\rm s}^4(n_{\rm l}^2-1)-2n_{\rm s}^2-3n_{\rm l}^2+1\bigg]\nonumber\\
a_\perp\;&=\;&\frac{1}{n_{\rm l}^2}\frac{n_{\rm l}^2-n_{\rm s}^2}{(n_{\rm s}^2-1)^2(n_{\rm s}^2+1)}
\bigg[n_{\rm s}^4(4n_{\rm s}^2-3n_{\rm s}-3)-n_{\rm s}^2(12n_{\rm s}^6-6n_{\rm s}^5+2)(n_{\rm l}^2-1)+n_{\rm l}^2(2n_{\rm s}^6+7n_{\rm s}^4-3n_{\rm s}^3+2)\bigg]\nonumber\\
\;&+\;&\frac{n_{\rm s}^2}{n_{\rm l}^2}\frac{n_{\rm l}^2-n_{\rm s}^2}{(n_{\rm s}^2-1)^{5/2}}\ln\left(\sqrt{n_{\rm s}^2-1}+n_{\rm s}\right)
\bigg[ n_{\rm l}^2(4n_{\rm s}^6-6n_{\rm s}^4+3n_{\rm s}^2-1)-n_{\rm s}^2(2n_{\rm s}^2-1)^2 \bigg]\nonumber\\
\;&+\;&\frac{n_{\rm s}^6}{2n_{\rm l}^2}\frac{n_{\rm l}^2-n_{\rm s}^2}{(n_{\rm s}^2-1)^2(n_{\rm s}^2+1)^{3/2}}
\ln\left(\frac{\sqrt{n_{\rm s}^2+1}+1}{\sqrt{n_{\rm s}^2+1}-1}\frac{\sqrt{n_{\rm s}^2+1}-n_{\rm s}}{\sqrt{n_{\rm s}^2+1}+n_{\rm s}}\right)
\bigg[4n_{\rm s}^4(n_{\rm l}^2-1)+2n_{\rm s}^2(n_{\rm l}^2-2)-3n_{\rm l}^2+1\bigg]\nonumber
\end{eqnarray}
\end{widetext}
Both, $a_\parallel$ and $a_\perp$, are positive for $n_{\rm l}>n_{\rm s}$ so that, as one would expect, the interaction, as compared to a half-space alone, is enhanced by the thin dielectric layer of the higher refractive index $n_{\rm l}$. The above result simplifies significantly in the case when $n_{\rm s}$ approaches unity i.e. when the situation resembles that of an atom interacting with a dielectric slab of refractive index $n_{\rm l}$. The coefficients $a_\parallel$ and $a_\perp$ reduce then to those recently calculated in \cite{slab} and are given by
\begin{eqnarray}
a_\parallel\;&=\;&\frac{(n_{\rm l}^2-1)(9n_{\rm l}^2+5)}{10n_{\rm l}^2}\;,\;\nonumber\\
a_\perp\;&=\;&\frac{(n_{\rm l}^2-1)(5n_{\rm l}^2+4)}{10n_{\rm l}^2}\;.\nonumber
\end{eqnarray}

\subsubsection{Thick layer ($\z/L \ll  1$)}\label{LargeLRetarded}
Here we assume that the thickness of the top layer is much greater than the distance between the atom and the surface, but which is still large enough for retardation to occur. Note that the reflection coefficient $\tilde{R}_\lambda^R$ (\ref{R}) can be separated into $L$-dependent and $L$-independent parts in the following manner:
\begin{equation}
\tilde{R}^R_\lambda=r_\lambda^{\rm vl}+ \frac{[1-(r_\lambda^{\rm vl})^2]r_\lambda^{\rm ls}e^{2i\kzs L}}{1+r_\lambda^{\rm vl}r_\lambda^{\rm ls}e^{2iL\kzs L}}.\label{RCoeffAlt}
\end{equation}
This way of writing the reflection coefficient splits the energy shift (\ref{RetAsym}) into a shift due to the single interface of refractive index $n_{\rm l}$ and corrections due to the finite thickness and the underlying material. It can be shown numerically, see Sec. \ref{sec:Num}, that for large values of $L$ the correction term is vanishingly small and can be safely discarded. Brute-force asymptotic analysis allows us to draw similar conclusions as in the electrostatic case, Section \ref{Sect:El:smalld}. To leading order the interaction gets altered by the same amount regardless of the position of the atom with respect to the interface. The next-to-leading-order correction is proportional to $\z/L^5$.

\subsection{Excited atoms. Non-retarded limit, ($2\z |E_{ji}|\ll 1$)}
The energy shift of an excited atom is given by equations
(\ref{Excited1}) and (\ref{Excited2}). The "non-resonant" part,
i.e. Eq. (\ref{Excited1}) has the same form as the energy shift of the
ground state atom and has been analysed in the previous
section. Therefore we now focus on the "resonant" part of the interaction that is given by equation (\ref{Excited2}). In order to conveniently obtain the non-retarded limit of (\ref{Excited2}) we will work with its slightly modified form given in equations (\ref{ExcitedElstatic1}) and (\ref{ExcitedElstatic2}). 

We start by noting that close to the interface we expect asymptotic
series to be in the inverse powers of $\z$. Equation
(\ref{ExcitedElstatic1}), where the $\eta$ integration runs over
$\eta\in[0,1]$, contributes only positive powers of $\z$. This is most
easily seen by expanding the exponential $\exp(2i|E_{ji}|\z\eta)$ about
origin as we may do in the limit $2\z|E_{ji}|\rightarrow 0$. Therefore,
to leading-order in the electrostatic limit, only
(\ref{ExcitedElstatic2}) contributes. Further we analyse
(\ref{ExcitedElstatic2}) by setting $\eta=\beta/(|E_{ji}|\z)$. Then,
according to (\ref{keta}), in the limit $|E_{ji}|\z\rightarrow 0$ the
wave vectors can effectively be approximated as
\begin{equation}
\kz\approx\kzs\approx\kzd\approx i \frac\beta\z\;.
\end{equation}
Then the result for the energy shift, after substituting
$\beta=k\z$, reduces to 
\begin{eqnarray}
\Delta E^{\rm res, el}=-\frac{1}{8\pi\epsilon_0}\sum_{j<i}\left(|\mu_\parallel|^2+2|\mu_\perp|^2\right)\hspace{- 1mm}\int_0^\infty\rd k k^2e^{-2k\z}\nonumber\\
\times\dfrac{\dfrac{n_{\rm l}^2-1}{n_{\rm l}^2+1}-\dfrac{n_{\rm l}^2-n_{\rm s}^2}{n_{\rm s}^2+n_{\rm l}^2}e^{-2kL}}{1-\dfrac{n_{\rm l}^2-1}{n_{\rm l}^2+1}\dfrac{n_{\rm l}^2-n_{\rm s}^2}{n_{\rm s}^2+n_{\rm l}^2}e^{-2kL}}.\hspace{2 mm}\label{elstaticExcited}
\end{eqnarray}
This result turns out to have the same dependence on $\z$ and $L$ as the
Coulomb interaction of the ground state atom, cf. Eq. (\ref{elstatic});
therefore we shall not analyse Eq. (\ref{elstaticExcited}) any
further. Note however, that the dependence on the atomic states is
different in equations (\ref{elstatic}) and (\ref{elstaticExcited}). We
would also like to point out that in the electrostatic limit, \emph{to
the order we are considering}, the quantity $\Delta E^{\rm res, el}$
turns out to be real, which would imply that the corrections to the
decay rates vanish. However, this conclusion is incorrect as it is known
that the change of spontaneous emission in the non-retarded limit is in
fact constant for a non-dispersive dielectric half-space
\cite{Wu}. However, any serious analysis of the changes of the decay
rates induced by a surface needs to take into account the absorption of
the material, which in the non-retarded limit plays a crucial role and
cannot be neglected. Furthermore we note that we have started from
Eq. (\ref{Excited2}), which, as explained before, contains poles on the
real axis signalling the trapped modes. However, the denominator of
(\ref{elstaticExcited}) never vanishes which reflects the fact that in
the electrostatic limit the trapped modes cease to exist and do not
contribute towards the energy shifts, as first mentioned in \cite{sipe}.

\subsection{Excited atoms. Retarded limit, ($2\z |E_{ji}|\gg 1$)}\label{sec:ExcitedAsym}
The leading-order behaviour of equation (\ref{Excited2}) in the retarded limit can be obtained by repeated integration by parts. Unlike in the electrostatic case now both equations, Eq. (\ref{ExcitedElstatic1}) and Eq. (\ref{ExcitedElstatic2}) contribute. We integrate them by parts and note that the non-oscillatory contributions that arise from the boundary terms evaluated at $\eta=0$ cancel out. It turns out that the leading-order contributions to the energy shift are due to the perpendicular component of the atomic dipole moment. They dominate the retarded interaction energy and behave as $\z^{-1}$. The contributions due to the component of the atomic dipole moment that is perpendicular to the surface contribute only terms proportional to $\z^{-2}$. We find that in the retarded limit the interaction energy up to the leading-order is given by
\begin{eqnarray}
\Delta E_i^{\rm res,ret}=
-\frac{1}{8\pi\epsilon_0\z}\sum_{j<i}|E_{ji}|^2|\mu_\parallel|^2\hspace{2.5 cm}
\nonumber\\
\times\frac{1}{1+2r_{\rm vl}r_{\rm ls}\cos(2|Eji|\tau)+r^2_{\rm vl}r^2_{\rm ls}}\hspace{2.7 cm}\nonumber\\
\times\left\{r_{\rm vl}(1+r^2_{\rm ls})\cos(2|E_{ji}|\z)\right.\hspace{3 cm}\nonumber\\\left.+r^2_{\rm vl}r_{\rm ls}\cos[2|E_{ji}|(\z-\tau)]\right.\hspace{1.75 cm}\nonumber\\
\left.+r_{\rm ls}\cos[2|E_{ji}|(\z+\tau)]\right\},\hspace{.5 cm}\label{retExcited}
\end{eqnarray}
where we have defined the optical thickness of the layer as $\tau=n_{\rm l}L$ and
\begin{equation}
r_{\rm vl}=\frac{1-n_{\rm l}}{1+n_{\rm l}},\;\;\;r_{\rm ls}=\frac{n_{\rm l}-n_{\rm s}}{n_{\rm l}+n_{\rm s}}.
\end{equation}
The final result agrees with that derived for a half-space in \cite{Wu}
if we take either $L\rightarrow 0$ or $n_{\rm l}\rightarrow n_{\rm s}$,
which is a consistency check of our calculation. However, the limit of
perfect reflectivity of the top layer does not make sense and one has to
start from equation (\ref{Excited2}) and rewrite the reflection
coefficient in the form (\ref{RCoeffAlt}) in order to study this case. 

Equation (\ref{retExcited}) is valid only approximately when the
distance between the atom and the surface is much greater than the
wavelength of the strongest atomic dipole transition, but it
nevertheless allows us to draw important conclusions. We note that the
interaction is resonant i.e. it is enhanced for certain values of
$LE_{ji}$. The most convenient way to understand the essence of these
resonance effects is to take the slab limit of equation
(\ref{retExcited}) i.e. set $n_{\rm s}=1$. In this limit we have
\begin{eqnarray}
\Delta E_i^{\rm res,ret}=
-\frac{1}{8\pi\epsilon_0\z}\sum_{j<i}|E_{ji}|^2|\mu_\parallel|^2\hspace{2.5 cm}\nonumber\\
\times\frac{1}{1-2r^2_{\rm vl}\cos(2|Eji|\tau)+r^4_{\rm vl}}\hspace{3.4 cm}\nonumber\\
\times\left\{r_{\rm vl}(1+r^2_{\rm vl})\cos(2|E_{ji}|\z)\right.\hspace{3 cm}\nonumber\\\left.-r^3_{\rm vl}\cos[2|E_{ji}|(\z-\tau)]\right.\hspace{1.75 cm}\nonumber\\
\left.-r_{\rm vl}\cos[2|E_{ji}|(\z+\tau)]\right\}.\hspace{.5 cm}\label{retExcitedSlab}
\end{eqnarray}
It is easily seen that whenever ${\cos(2|Eji|\tau)=1}$ then $\Delta
E_i^{\rm res,ret}=0$, i.e. the leading-order interaction vanishes.
Conversely, the amplitude of oscillations in equation
(\ref{retExcitedSlab}) is maximized when
$\cos(2|Eji|\tau)=-1$. Therefore we have a condition for resonance in
terms of the wavelength of the strongest atomic dipole transition
$\lambda_{ji}$
\begin{equation}
\tau=nL=\frac{\lambda_{ji}}{2}\left(\kappa+\frac{1}{2}\right),\;\;\;\kappa=0,1,2\;\ldots\label{Resonance}
\end{equation}
Eq. (\ref{Resonance}) holds for $Z|E_{ji}|\gg 1$ but if the value of
$Z|E_{ji}|$ approaches unity, the relation loses its validity, because complications arise from the fact that when the atom is close to the surface the evanescent waves come into play whereas the condition ($\ref{Resonance}$) refers to the interaction of an atom with travelling modes only. In the non-retarded limit $\z|E_{ji}|\ll 1$ the notion of resonance loses its meaning altogether, cf. Eq. (\ref{elstaticExcited}). Exploring the extreme case in the retarded limit we note that at anti-resonance i.e. when 
\begin{equation}
\tau=nL=\frac{\lambda_{ji}}{2}\kappa,\;\;\;\kappa=0,1,2\;\ldots\label{AntiResonance}
\end{equation}
equation (\ref{retExcited}) becomes
\begin{equation}
\Delta E_i^{\rm res,ret}=
\frac{1}{8\pi\epsilon_0\z}\frac{n_{\rm s}-1}{n_{\rm s}+1}\sum_{j<i}|E_{ji}|^2|\mu_\parallel|^2\cos(2|E_{ji}|\z),\;\;\label{AntiResonanceShift}
\end{equation}
i.e. the atom does not feel the presence of the layer and the
interaction assumes the form of that between an atom and a single
half-space of refractive index $n_{\rm s}$, cf. \cite{Wu}. This means
that in the retarded regime the leading-order interaction between an
excited atom and a slab of thickness $L$ vanishes whenever the optical
thickness of the slab $\tau=n_{\rm l}L$ is equal to a half-integer
multiple of the wavelength of the dominant atomic transition
$\lambda_{ji}$ (cf. also Fig. \ref{fig:Excited2} later on). Conversely, at resonance the shift becomes
\begin{equation}
\Delta E_i^{\rm res,ret}=
\frac{1}{8\pi\epsilon_0\z}\frac{n^2_{\rm l}-n_{\rm s}}{n^2_{\rm l}+n_{\rm s}}\sum_{j<i}|E_{ji}|^2|\mu_\parallel|^2\cos(2|E_{ji}|\z),\;\;\label{ResonanceShift}
\end{equation}
so that the amplitude of oscillations exceeds the amplitude that would have been caused by a single half-space of refractive index $n_{\rm l}$. It also reaches the perfect reflector limit $n_{\rm l}\rightarrow \infty$ more rapidly. Finally, we shall also remark that the meaning of the conditions (\ref{Resonance}) and (\ref{AntiResonance}) is interchanged if the refractive index of the substrate $n_{\rm s}$ exceeds that of the layer $n_{\rm l}$ i.e. when $n_{\rm s}>n_{\rm l}$.

\section{Numerical Examples}\label{sec:Num}
In this section we present a few numerical results designed to illustrate the influence of the dielectric layer on the Casimir-Polder interaction between an atom and a dielectric half-space. In practice, the sum over intermediate states $j$ in Eq. (\ref{RetAsym}) and  in Eq. (\ref{Excited2}) is restricted to one or a few states to which there are strong dipole transitions. Hence, we assume a two-level system in which $E_{ji}$ is a single number, namely the energy spacing of the levels with the strongest dipole transition. Additionally, we focus just on the contributions to the energy shift due to the component of the atomic dipole that is parallel to the interface of the dielectrics. The contributions due to the perpendicular components of the atomic dipole moment can be easily generated with from Eq. (\ref{RetAsym}) using standard computer algebra packages like Mathematica or Maple. We start by simple checks on the asymptotic expansions derived in the previous section. 
\subsection{Ground-state atoms}
\begin{figure}[ht]
\hspace{-8mm}
\includegraphics[width=9.0 cm, height=7.0 cm]{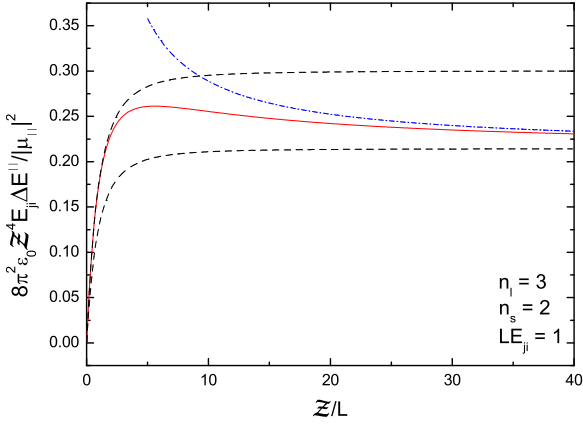}
\caption{\label{fig:numerics1}Plot of the exact energy-level shift contributions $\Delta E^{\parallel}$ (solid), Eq. (\ref{RetAsym}), multiplied by $\z^4$. Dashed lines represent the energy shifts due to the single dielectric half-spaces of refractive indices $n_{\rm l}$ (top) and $n_{\rm s}$ (bottom), whereas the dotted-dashed lines represents the asymptotic approximation (\ref{LargeDAsym}).}
\end{figure}
\begin{figure}[ht]
\hspace{-8mm}
\includegraphics[width=9.0 cm, height=7.0 cm]{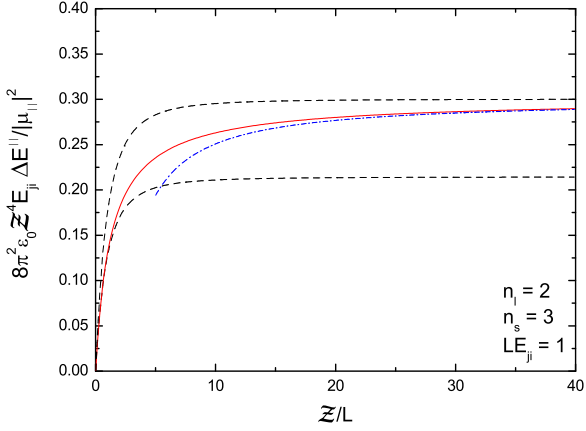}
\caption{\label{fig:numerics2}Plot of the exact energy-level shift $\Delta E^\parallel$ (solid), Eq. (\ref{RetAsym}), multiplied by $\z^4$. Dashed lines represent the energy shifts due to the single dielectric half-spaces of refractive indices $n_{\rm l}$ (bottom) and $n_{\rm s}$ (top), whereas the dotted-dashed lines represents the asymptotic approximation (\ref{LargeDAsym}).}
\end{figure}
We choose to plot the energy-level shift $\Delta E$ multiplied by $\z^4$
so that the asymptotic behaviour of it as a function of distance is more
apparent,  because $\z^4\Delta E$ for a dielectric half-space approaches
constant \cite{Wu}. Then, one can easily track the variation of the
energy shift caused by the top layer as compared to the half-space
shifts, Fig. \ref{fig:numerics1} and Fig. \ref{fig:numerics2}. We remark
that even though the derivation of the energy shift in this paper was
based on the assumption $n_{\rm l}>n_{\rm s}$, the results are also
valid in the case when the top layer has a smaller reflectivity than the
substrate. In such a case the result can be used e.g. to model a thin
layer of oxide or any kind of dirt on the substrate which is often
present under realistic conditions. 

\begin{figure}[ht]
\hspace{-8mm}
\includegraphics[width=9.0 cm, height=7.0 cm]{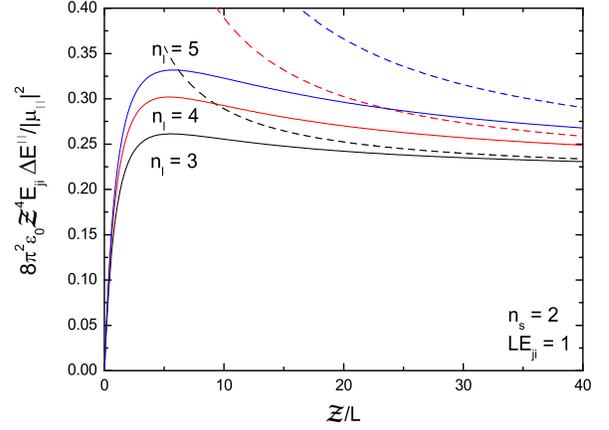}
\caption{\label{fig:numerics3}Plot of the exact energy shift $\Delta E^\parallel$, (solid, Eq. (\ref{RetAsym})), multiplied by $\z^4$ together with the asymptotic approximations (dashed, Eq. (\ref{LargeDAsym})).}
\end{figure}

The asymptotic expansion (\ref{LargeDAsym}) works well for large $\z/L$
and not too high values of the refractive index $n_{\rm l}$. This is
demonstrated in Fig. \ref{fig:numerics3}. The increase of the refractive
index $n_{\rm l}$ has an impact on the accuracy of the approximation which is valid provided
\begin{equation}
\z\gg \lambda_{ji}+\tau_{\rm l}
\end{equation} 
with $\lambda_{ji}$ being the wavelength of the dominant atomic transition and $\tau_{\rm l}=n_{\rm l}L$ is the optical thickness of the top layer.
\begin{figure}[ht]
\hspace{-8mm}
\includegraphics[width=9.0 cm, height=7.0 cm]{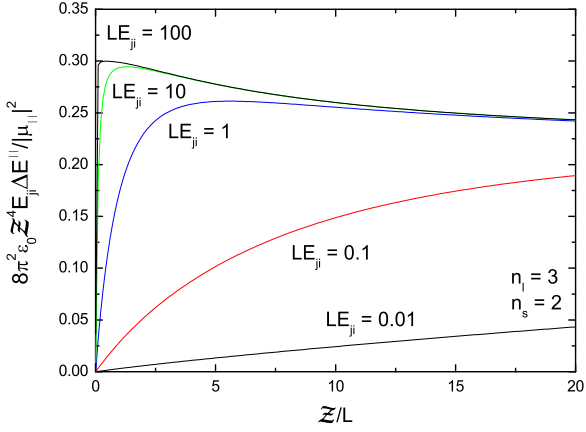}
\caption{\label{fig:numerics4}Plot of the exact energy shift $\Delta E^\parallel$ (Eq. (\ref{RetAsym})) multiplied by $\z^4$ as a function of $\z/L$ for various values of the retardation parameter $E_{ji}L$.}
\end{figure}
\begin{figure}[ht]
\hspace{-8mm}
\includegraphics[width=9.0 cm, height=7.0 cm]{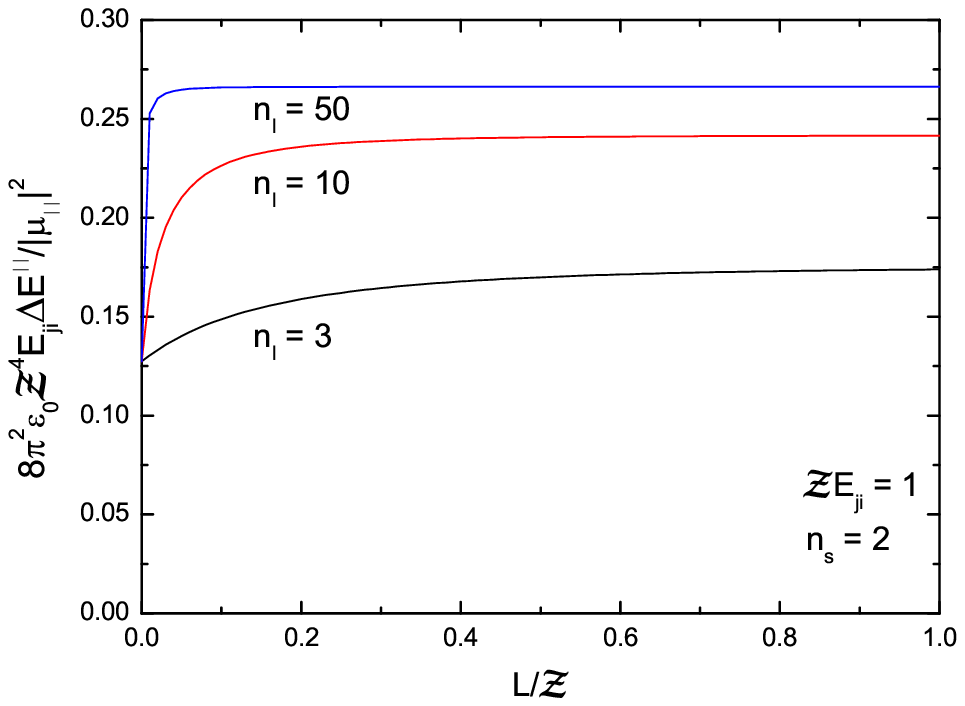}
\caption{\label{fig:numerics5}Plot of the exact energy shift $\Delta E^\parallel$ (Eq. (\ref{RetAsym})) multiplied by $\z^4$ as a function of layer's thickness $L$ measured in units of fixed atom-wall separation $\z$ for various values of the layer's refractive index $n_{\rm l}>n_{\rm s}$.}
\end{figure}
\begin{figure}[ht]
\hspace{-8mm}
\includegraphics[width=9.0 cm, height=7.0 cm]{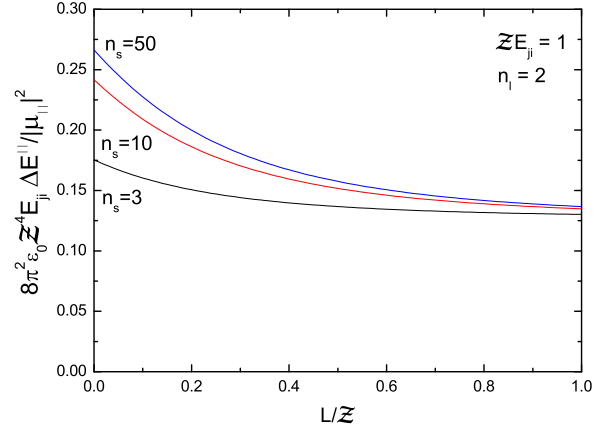}
\caption{\label{fig:numerics6}Plot of the exact energy shift $\Delta E^\parallel$ (Eq. (\ref{RetAsym})) multiplied by $\z^4$ as a function of layer's thickness $L$ measured in units of fixed atom-wall separation $\z$ for various values of the substrate's refractive index $n_{\rm s}>n_{\rm l}$.}
\end{figure}
In Fig. \ref{fig:numerics4} we demonstrate the behaviour of the energy shift depending on the various values of the parameter $E_{ji}$ measured in units of the layer's thickness. For small $E_{ji}$ we clearly observe linear behaviour that corresponds to the $\z^{-3}$ dependence of the shift in the electrostatic regime.

We also find it instructive to plot the energy-level shift as a function of the thickness of the top layer $L$ for different values of the refractive index $n_{\rm l}$ while keeping the distance of the atom from the surface fixed, Fig. \ref{fig:numerics5} and {Fig. \ref{fig:numerics6}}.

\subsection{Excited atoms}
The energy shift of an excited atom splits into two distinct parts,
cf. Eq. (\ref{Excited1}) and Eq. (\ref{Excited2}). The non-oscillatory
part displays the same behaviour as the energy shift of the ground-state
atoms, which we have already analysed numerically in the previous
section. Here we will focus on the oscillatory contributions to the
level shifts that are given by Eq. (\ref{Excited2}). We choose to plot
the dimensionless integrals contained in equations
(\ref{ExcitedElstatic1}) and (\ref{ExcitedElstatic2}) as this is
numerically more efficient than plotting the integral in
Eq. (\ref{Excited2}). It should be borne in mind that the reflection
coefficients contain the dispersion relation in denominators that now
has solutions on the real axis. For the purpose of the present
demonstration it is sufficient to simply displace the poles off the real
axis by adding small imaginary part to the denominator of the reflection
coefficients, which amounts to taking the Cauchy principal-value during numerical integration.
\begin{figure}[ht]
\hspace{-8mm}
\includegraphics[width=9.0 cm, height=7.0 cm]{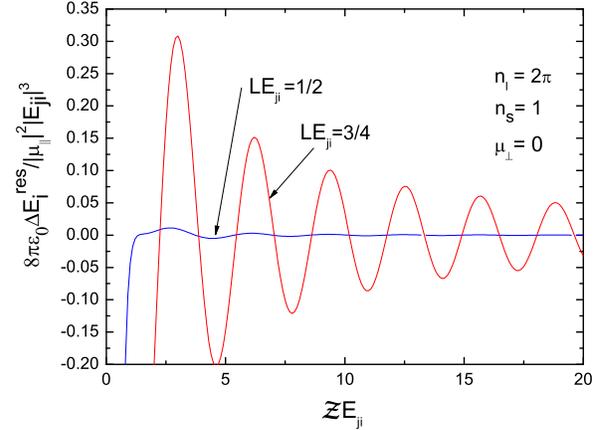}
\caption{\label{fig:Excited2}Plot of the exact energy-level shift (\ref{Excited2}) (resonant part) in an excited atom due to the parallel component of the atomic dipole moment placed in front of a slab of thickness $L$ and refractive index $n_{\rm l}=2\pi$. The energy spacing of the dominant atomic transition is such that $LE_{ji}=3/4$ i.e. it satisfies the resonance condition (\ref{AntiResonance}). As is seen, when $LE_{ji}=1/2$, the energy shift in the retarded regime is strongly suppressed, cf. Eq. (\ref{retExcited}).}
\end{figure}
\begin{figure}[ht]
\hspace{-8mm}
\includegraphics[width=9.0 cm, height=7.0 cm]{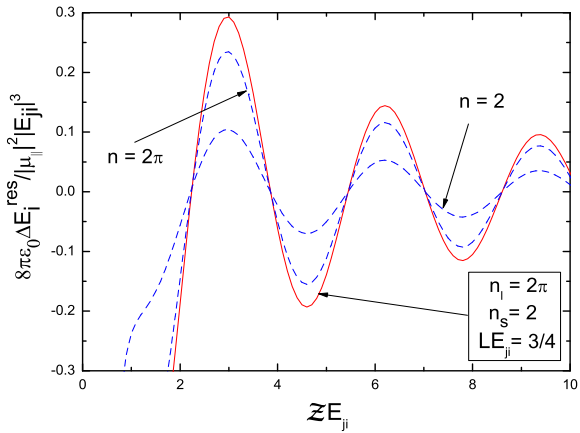}
\caption{\label{fig:Excited3}Plot of the exact energy-level shift (\ref{Excited2}) (resonant part) in an excited atom  due to the parallel component of  the atomic dipole moment placed in front of the layered dielectric with parameters as shown on the graph (solid).  The resonant condition (\ref{Resonance}) is satisfied  so that the interaction is enhanced. The amplitude of oscillations exceeds the one that would have been caused by an unlayered half-space of the refractive index $n=2\pi$, cf. Eq. (\ref{AntiResonanceShift}). Compare also Fig. \ref{fig:numerics1}. The dashed lines represent the interaction between an atom and single half-space of refractive index $n$ as indicated.}
\end{figure}
\begin{figure}[ht]
 \hspace{-8 mm}
    \includegraphics[width=9.0 cm, height=7.0 cm]{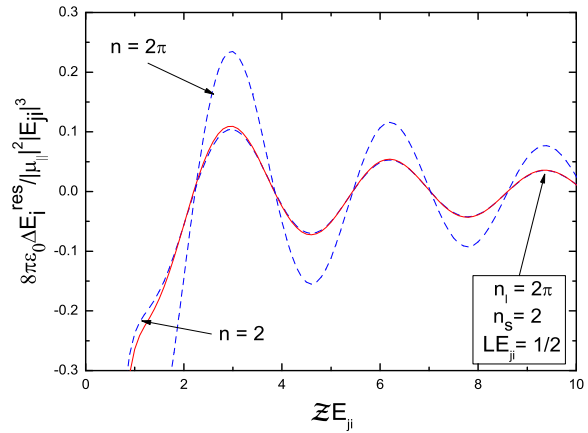}
  \caption{\label{fig:Excited4}Plot of the exact energy-level shift (\ref{Excited2}) (resonant part) in an excited atom  due to the parallel component of the atomic dipole moment placed in front of the layered dielectric with parameters as shown on the graph (solid). The anti-resonant condition (\ref{AntiResonance}) is satisfied  so that the presence of the layer is almost unnoticeable, cf. Eq. (\ref{AntiResonanceShift}). The dashed lines represent the interaction between an atom and single half-space of refractive index $n$ as indicated.}
\end{figure}
\begin{figure}[htbp]
  \hspace{-8 mm}
    \includegraphics[width=9.0 cm, height=7.0 cm]{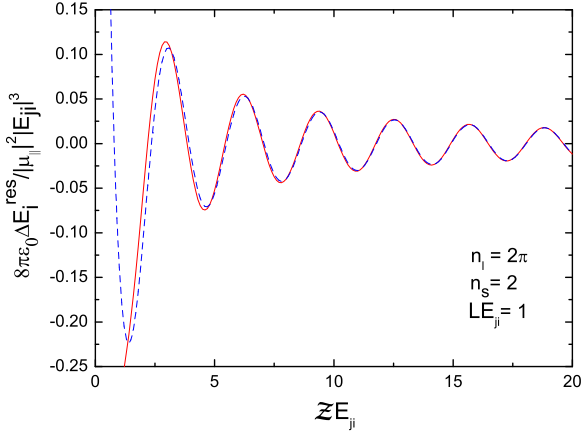}
  \caption{\label{fig:Excited5}Plot of the exact energy-level shift (\ref{Excited2}) (resonant part) in an excited atom  due to the parallel component of  the atomic dipole moment placed in front of the layered dielectric with parameters as shown on the graph (solid). The dashed line represents the approximation in the retarded regime, Eq. (\ref{retExcited}).}
\end{figure}
\begin{figure}[htbp]
  \hspace{-8 mm}
    \includegraphics[width=9.0 cm, height=7.0 cm]{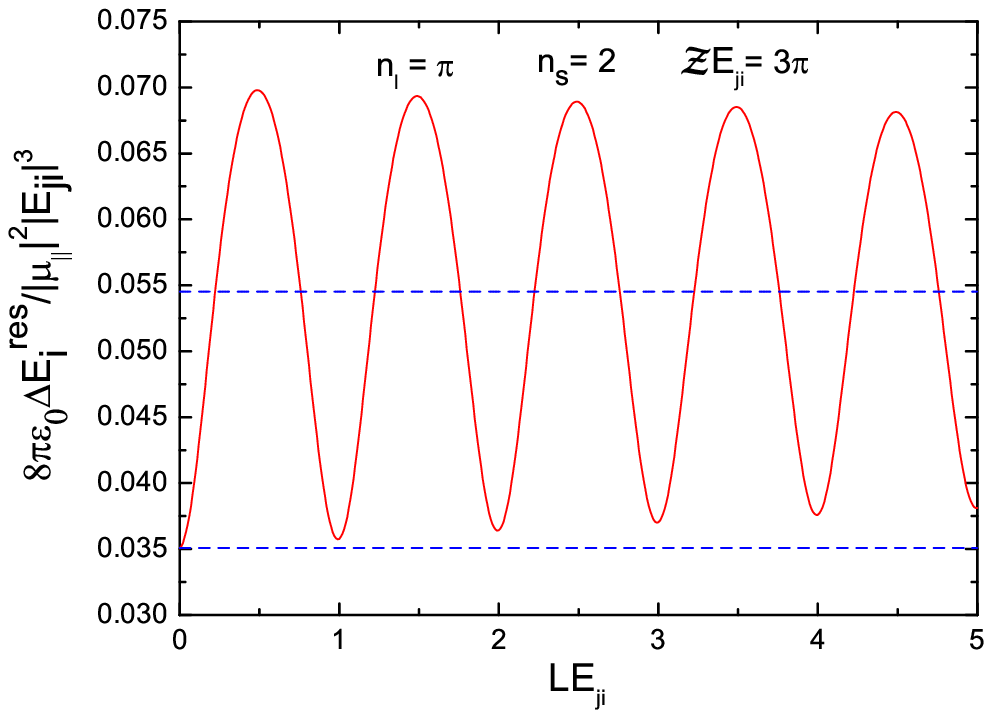}
  \caption{\label{fig:Excited6}Plot of the exact energy-level shift (\ref{Excited2}) (resonant part) in an excited atom  due to the parallel component of  the atomic dipole moment placed in front of the layered dielectric with parameters as shown on the graph (solid). The dashed lines represent energy shifts caused by the single half-spaces of refractive index $n_{\rm l}=2\pi$ (top) and $n_{\rm s}=2$ (bottom).}
\end{figure}
In Fig. \ref{fig:Excited2} we demonstrate that indeed, if the anti-resonance condition (\ref{AntiResonance}) is satisfied, the interaction energy between the excited atom and the slab is strongly suppressed for $\z E_{ji}\gg 1$. In general, for the layered dielectric rather than the slab, the effect of resonance is shown in Fig. \ref{fig:Excited3} and Fig. \ref{fig:Excited4}. Note that the energy-level shift in an excited atom due to the layered dielectric can be significantly enhanced. Unlike in the case of the ground state atom where the energy shift caused by the layered structure of refractive indices $n_{\rm l}$ and $n_{\rm s}$ is bounded by the single half-space shifts (compare Fig. \ref{fig:numerics1}), the excited atom can experience shifts greater than those caused by the unlayered half-space of the refractive index $n={\rm max}(n_{\rm l},n_{\rm s})$, Fig. \ref{fig:Excited3}, which is due to resonance effects. Conversely, it is also possible that the interaction with the layer will be unnoticeable if the anti-resonance condition (\ref{AntiResonance}) is satisfied, Fig. \ref{fig:Excited4}. Next, in Fig. \ref{fig:Excited5}, we show that the approximation of Eq. (\ref{Excited2}) derived in (\ref{retExcited}) turns out to be quite accurate and can be safely used to quickly estimate the energy shift in an excited atom caused by the layered dielectric, provided the condition $\z E_{ji}\gg 1$ is satisfied. It is also interesting to plot the resonant part of the energy shift as a function of $LE_{ji}$ while keeping $ZE_{ji}$ fixed. This is done in Fig. \ref{fig:Excited6}. It is seen that the energy shift indeed experiences the oscillatory resonant behaviour. The subsequent minima and maxima are less and less pronounced as the value of $LE_{ji}$ increases. This is because as we increase $LE_{ji}$ the resonances and anti-resonances move closer and closer together so that their effects cancel out. It is interesting to note that this behaviour could not have been inferred from equation (\ref{retExcited}), which indicates that the approximation (\ref{retExcited}) can be useful only for $LE_{ji}\ll 1$, which can also be easily verified numerically.

\section{Summary}
Using perturbation theory we have calculated the energy-level shift in a
neutral atom placed in front of a layered dielectric half-space, as
shown in Fig. \ref{fig:slab}. The major difficulty in working out the
energy shift is the sum over all modes that appears in this type of
calculation, Eq. (\ref{shift1}), especially when the spectrum of the
modes consists of the continuous and discrete parts, Sec. \ref{sec:trav}
and \ref{sec:trap}. This obstacle can be circumvented by using
complex-variable techniques to express the sum over all modes as a
single contour integral in the complex $\kz$-plane, Eq. (\ref{shift2})
and Fig. \ref{fig:c}. Then, the energy shift (\ref{RetAsym}) is easily
analyzed asymptotically as well as numerically. For a ground-state atom,
regardless of whether in retarded or non-retarded regimes, we find that
the leading-order correction to the interaction of an atom with an
unlayered interface is proportional to $L/\z$. The asymptotic series are
given by (\ref{ElstaticLargeD}) and (\ref{LargeDAsym}) and provide
reasonable estimate of the influence of the single dielectric layer on
the standard half-space result, Fig. \ref{fig:numerics3}. In the
opposite case of a very thick layer i.e. $\z/L <<1$ we find that the
result is well approximated by a dielectric half-space \cite{Wu}. For
excited atoms we find that the interaction between an atom and the
layered dielectric (\ref{Excited2}) is subject to resonances that occur
between the wavelength of the dominant atomic transition $\lambda_{ji}$
and the thickness of the layer $L$, Sec. \ref{sec:ExcitedAsym}. In
particular, the interaction between an atom and the slab can be strongly
suppressed in the retarded regime, cf. Fig. \ref{fig:Excited2}, whenever
the optical thickness of the slab $\tau$ is equal to the half-integer
multiple of the wavelength of the dominant atomic transition
$\lambda_{ji}$. The existence of resonance effects suggests a physical
picture of the excited atom as a radiating dipole. The resonance and
anti-resonance correspond to constructive and destructive interference.
We have also provided reasonable approximations in the non-retarded
(\ref{elstaticExcited}) and retarded (\ref{retExcited}) regimes that can
be used to quickly estimate the magnitude of the resonant interaction
between an atom and a layered dielectric.

\appendix

\section{Fresnel coefficients for layered dielectric}
\label{App:Fresnels}
Here we list the reflection and transmission coefficients appearing in the normal-modes of the system as discussed in the section \ref{sec:trav}. For the left-incident modes we find
\begin{eqnarray}
R_\lambda^L\;&=\;&\frac{r_\lambda^{\rm sl}+r_\lambda^{\rm lv}e^{2ik_{\rm zl}L}}{1+r_\lambda^{\rm sl}r_\lambda^{\rm lv}e^{2ik_{\rm zl}L}}e^{-ik_{\rm zs}L}\nonumber\\
I_\lambda^L\;&=\;&\frac{t_\lambda^{\rm sl}e^{i(k_{\rm zl}-k_{\rm zs})L/2}}{1+r_\lambda^{\rm sl}r_\lambda^{\rm lv}e^{2ik_{\rm zl}L}}\nonumber\\
J_\lambda^L\;&=\;&\frac{t_\lambda^{\rm sl}r_\lambda^{\rm lv}e^{(3ik_{\rm zl}-ik_{\rm zs})L/2}}{1+r_\lambda^{\rm sl}r_\lambda^{\rm lv}e^{2ik_{\rm zl}L}}\nonumber\\
T_\lambda^L\;&=\;&\frac{t_\lambda^{\rm sl}t_\lambda^{\rm lv}e^{(2ik_{\rm zl}-ik_{\rm zs}-ik_{\rm z})L/2}}{1+r_\lambda^{\rm sl}r_\lambda^{\rm lv}e^{2ik_{\rm zl}L}}\nonumber,
\end{eqnarray}
and for the right-incident modes we get
\begin{eqnarray}
R_\lambda^R\;&=\;&\frac{r_\lambda^{\rm vl}+r_\lambda^{\rm ls}e^{2ik_{\rm zl}L}}{1+r_\lambda^{\rm vl}r_\lambda^{\rm ls}e^{2ik_{\rm zl}L}}e^{-ik_{\rm z}L}\nonumber\\
I_\lambda^R\;&=\;&\frac{t_\lambda^{\rm vl}e^{i(k_{\rm zl}-k_{\rm z})L/2}}{1+r_\lambda^{\rm vl}r_\lambda^{\rm ls}e^{2ik_{\rm zl}L}}\nonumber\\
J_\lambda^R\;&=\;&\frac{t_\lambda^{\rm vl}r_\lambda^{\rm ls}e^{(3ik_{\rm zl}-ik_{\rm z})L/2}}{1+r_\lambda^{\rm vl}r_\lambda^{\rm ls}e^{2ik_{\rm zl}L}}\nonumber\\
T_\lambda^R\;&=\;&\frac{t_\lambda^{\rm vl}t_\lambda^{\rm ls}e^{(2ik_{\rm zl}-ik_{\rm zs}-ik_{\rm z})L/2}}{1+r_\lambda^{\rm vl}r_\lambda^{\rm ls}e^{2ik_{\rm zl}L}}.\nonumber
\end{eqnarray}
The Fresnel reflection coefficients $r_\lambda ^{\rm ab}$ for a single
interface are given by (\ref{Fresnel}).

\section{Electrostatic calculation of the energy-level shift in a ground-state atom in a layered geometry}\label{App:Elstatic}
To provide an additional check on the consistency of our calculations we would like to derive equation (\ref{elstatic}) by means of electrostatics. We start from the general formula derived in \cite{nonret} that expresses the electrostatic interaction energy of a electric dipole in the presence of a dielectric in terms of Green's function of the Laplace equation
\begin{equation}
\Delta E=\frac{1}{2}\sum_i\langle\mu_i^2\rangle\nabla_i\nabla'_i\;G_H(\br,\br')\bigg|_{\br=\br_0,\br'=\br_0}.\label{DipoleEnergy}
\end{equation}
Here the sum runs over three components of the dipole moment and the subscript $H$ means that only the homogeneous correction to the free-space Green's function that is caused by the presence of the boundary enters the formula. This ensures that the self-energy of the dipole is omitted and guarantees the convergence of the final result. The harmonic function $G_H(\br,\br')$ is a solution of the Laplace equation that vanishes for $|z|\rightarrow \infty$. Therefore it can be written in the form:
\begin{eqnarray}
G_H(\br,\br')=-\frac{1}{4\pi\epsilon_0}\int_0^\infty\rd_2 \bkp e^{i\bkp\cdot\brp}\hspace{3 cm}\nonumber\\
\times\left\{
\begin{array}{lc}
C_1(\bkp,\br')e^{k_z z}& z<L/2\\
C_2(\bkp,\br')e^{k_z z}+C_3(\bkp,\br')e^{-k_z z}&|z|<L/2\\
C_4(\bkp,\br')e^{-k_z z}& z>L/2
\end{array}\right. ,\;\;\;
\end{eqnarray}
with $k_z=\sqrt{k_x^2+k_y^2}$. The $C$ coefficients are easily worked out by applying the continuity conditions, which result from Maxwell's equations, across the interfaces and one finds that
\begin{eqnarray}
G_H(\br,\br')=-\frac{1}{4\pi\epsilon_0}\int_0^\infty\rd k J_0(k\rho)e^{-k(z+z')}\hspace{1 cm}\nonumber\\
\times \dfrac{\dfrac{n_{\rm l}^2-1}{n_{\rm l}^2+1}-\dfrac{n_{\rm l}^2-n_{\rm s}^2}{n_{\rm s}^2+n_{\rm l}^2}e^{-2kL}}{1-\dfrac{n_{\rm l}^2-1}{n_{\rm l}^2+1}\dfrac{n_{\rm l}^2-n_{\rm s}^2}{n_{\rm s}^2+n_{\rm l}^2}e^{-2kL}}\;
\end{eqnarray}
with $\rho=\sqrt{(x-x')^2+(y-y')^2}$. Application of the formula (\ref{DipoleEnergy}) is straightforward and we easily derive that the electrostatic interaction energy of a dipole in a vicinity of the layered dielectric is indeed equal to (\ref{elstatic}).

\section{Retarded limit of the interaction energy between an atom and a dielectric half-space}\label{App:HSResult}
The interaction between an atom and a non-dispersive dielectric half-space has been considered in detail in \cite{Wu}. It has been shown there that, to leading-order, the energy shift in the retarded limit can be expressed as
\begin{equation}
\Delta E_{n_{\rm s}}^{\rm ret}=-\frac{3}{64\pi^2\epsilon_0\z^4}\sum_{j\neq i}
\left(\frac{c_\parallel|\mu_\parallel|^2+c_\perp|\mu_\perp|^2}{E_{ji}}\right),
\end{equation}
with the coefficients $c_{\parallel,\perp}$ given by
\begin{eqnarray}
c_\parallel &=& -\frac{1}{n_{\rm s}^2-1}\left(\frac{2}{3}n_{\rm s}^2+n_{\rm s}-\frac{8}{3}\right)\nonumber\\
&+&\frac{2n_{\rm s}^4}{(n_{\rm s}^2-1)\sqrt{n_{\rm s}^2+1}}\ln\left(\frac{\sqrt{n_{\rm s}^2+1}+1}{n_{\rm s}\left[\sqrt{n_{\rm s}^2+1}+n_{\rm s}\right]}\right)\nonumber\\
&+&\frac{2n_{\rm s}^4-2n_{\rm s}^2-1}{(n_{\rm s}^2-1)^{3/2}}\ln\left(\sqrt{n_{\rm s}^2+1}+n_{\rm s}\right),\nonumber\\
c_\perp &=& \frac{1}{n_{\rm s}^2-1}\left(4n_{\rm s}^4-2n_{\rm s}^3-\frac{4}{3}n_{\rm s}^2+\frac{4}{3}\right)\nonumber\\
&-&\frac{4n_{\rm s}^6}{(n_{\rm s}^2-1)\sqrt{n_{\rm s}^2+1}}\ln\left(\frac{\sqrt{n_{\rm s}^2+1}+1}{n_{\rm s}\left[\sqrt{n_{\rm s}^2+1}+n_{\rm s}\right]}\right)\nonumber\\
&-&\frac{2n_{\rm s}^2(2n_{\rm s}^4-2n_{\rm s}^2+1)}{(n_{\rm s}^2-1)^{3/2}}\ln\left(\sqrt{n_{\rm s}^2-1}+n_{\rm s}\right).\nonumber
\end{eqnarray}

\end{document}